%


\newcount\mgnf\newcount\tipi\newcount\tipoformule\newcount\driver
\newcount\indice
\driver=1        
\mgnf=0          
\tipi=2          
\tipoformule=0   
\indice=1        

\ifnum\mgnf=0
   \magnification=\magstep0
   \hsize=14truecm\vsize=24.truecm
   \parindent=0.3cm\baselineskip=0.45cm\fi
\ifnum\mgnf=1
   \magnification=\magstep1\hoffset=0.truecm
   \hsize=14truecm\vsize=24.truecm
   \baselineskip=18truept plus0.1pt minus0.1pt \parindent=0.9truecm
   \lineskip=0.5truecm\lineskiplimit=0.1pt      \parskip=0.1pt plus1pt\fi

\let\a=\alpha \let\b=\beta  \let\g=\gamma     \let\d=\delta  \let\e=\varepsilon
\let\z=\zeta  \let\h=\eta   \let\th=\vartheta \let\k=\kappa   \let\l=\lambda
\let\m=\mu    \let\n=\nu    \let\x=\xi        \let\p=\pi      \let\r=\rho
\let\s=\sigma \let\t=\tau   \let\f=\varphi     
               \let\o=\omega

\let\G=\Gamma \let\D=\Delta     \let\L=\Lambda  
     \let\F=\Phi      
\let\O=\Omega 


\global\newcount\numsec\global\newcount\numfor
\global\newcount\numapp\global\newcount\numcap
\global\newcount\numfig\global\newcount\numpag
\global\newcount\numnf

\def\SIA #1,#2,#3 {\senondefinito{#1#2}%
\expandafter\xdef\csname #1#2\endcsname{#3}\else
\write16{???? ma #1,#2 e' gia' stato definito !!!!} \fi}

\def \FU(#1)#2{\SIA fu,#1,#2 }
 
\def\etichetta(#1){(\veroparagrafo.\veraformula)%
\SIA e,#1,(\veroparagrafo.\veraformula) %
\global\advance\numfor by 1%
\write15{\string\FU (#1){\equ(#1)}}%
\write16{ EQ #1 ==> \equ(#1)  }}
\def\etichettaa(#1){(A\veraappendice.\veraformula)
 \SIA e,#1,(A\veraappendice.\veraformula)
 \global\advance\numfor by 1
 \write15{\string\FU (#1){\equ(#1)}}
 \write16{ EQ #1 ==> \equ(#1) }}
\def\getichetta(#1){Fig. \verafigura
 \SIA g,#1,{\verafigura}
 \global\advance\numfig by 1
 \write15{\string\FU (#1){\graf(#1)}}
 \write16{ Fig. #1 ==> \graf(#1) }}
\def\retichetta(#1){\numpag=\pgn\SIA r,#1,{\verapagina}
 \write15{\string\FU (#1){\rif(#1)}}
 \write16{\rif(#1) ha simbolo  #1  }}
\def\etichettan(#1){(n\verocapitolo.\veranformula)
 \SIA e,#1,(n\verocapitolo.\veranformula)
 \global\advance\numnf by 1
\write16{\equ(#1) <= #1  }}

\newdimen\gwidth
\gdef\profonditastruttura{\dp\strutbox}
\def\senondefinito#1{\expandafter\ifx\csname#1\endcsname\relax}
\def\BOZZA{
\def\alato(##1){
 {\vtop to \profonditastruttura{\baselineskip
 \profonditastruttura\vss
 \rlap{\kern-\hsize\kern-1.2truecm{$\scriptstyle##1$}}}}}
\def\galato(##1){ \gwidth=\hsize \divide\gwidth by 2
 {\vtop to \profonditastruttura{\baselineskip
 \profonditastruttura\vss
 \rlap{\kern-\gwidth\kern-1.2truecm{$\scriptstyle##1$}}}}}
\def\verapagina{
{\romannumeral\number\numcap}.\number\numsec.\number\numpag}}

\def\alato(#1){}
\def\galato(#1){}
\def\veroparagrafo{\number\numsec}\def\veraformula{\number\numfor}
\def\veraappendice{\number\numapp}
\def\verapagina{\number\pageno}\def\veranformula{\number\numnf}
\def\verafigura{{\romannumeral\number\numcap}.\number\numfig}
\def\verocapitolo{\number\numcap}\def\veranformula{\number\numnf}
\def\Eqn(#1){\eqno{\etichettan(#1)\alato(#1)}}
\def\eqn(#1){\etichettan(#1)\alato(#1)}

\def\Eq(#1){\eqno{\etichetta(#1)\alato(#1)}}
\def\eq(#1){\etichetta(#1)\alato(#1)}
\def\Eqa(#1){\eqno{\etichettaa(#1)\alato(#1)}}
\def\eqa(#1){\etichettaa(#1)\alato(#1)}
\def\dgraf(#1){\getichetta(#1)\galato(#1)}
\def\drif(#1){\retichetta(#1)}

\def\eqv(#1){\senondefinito{fu#1}$\clubsuit$#1\else\csname fu#1\endcsname\fi}
\def\equ(#1){\senondefinito{e#1}\eqv(#1)\else\csname e#1\endcsname\fi}
\def\graf(#1){\senondefinito{g#1}\eqv(#1)\else\csname g#1\endcsname\fi}
\def\rif(#1){\senondefinito{r#1}\eqv(#1)\else\csname r#1\endcsname\fi}

\openin14=\jobname.aux \ifeof14 \relax \else
\input \jobname.aux \closein14 \fi



\ifnum\tipoformule=1\let\Eq=\eqno\def\eq{}\let\Eqa=\eqno\def\eqa{}
\def\equ{}\fi


{\count255=\time\divide\count255 by 60 \xdef\hourmin{\number\count255}
	\multiply\count255 by-60\advance\count255 by\time
   \xdef\hourmin{\hourmin:\ifnum\count255<10 0\fi\the\count255}}

\def\oramin{\hourmin }

\def\data{\number\day/\ifcase\month\or gennaio \or febbraio \or marzo \or
aprile \or maggio \or giugno \or luglio \or agosto \or settembre
\or ottobre \or novembre \or dicembre \fi/\number\year;\ \oramin}


\newcount\pgn \pgn=1
\def\foglio{\number\numsec:\number\pgn
\global\advance\pgn by 1}
\def\foglioa{A\number\numsec:\number\pgn
\global\advance\pgn by 1}

%

\footline={\rlap{\hbox{\copy200}}}

%
%
\newdimen\xshift \newdimen\xwidth
%
%
\def\ins#1#2#3{\vbox to0pt{\kern-#2 \hbox{\kern#1 #3}\vss}\nointerlineskip}
%
%
\def\insertplot#1#2#3#4{
    \par \xwidth=#1 \xshift=\hsize \advance\xshift
     by-\xwidth \divide\xshift by 2 \vbox{
  \line{} \hbox{ \hskip\xshift  \vbox to #2{\vfil
 \ifnum\driver=0 #3  
                 \special{ps: plotfile #4.ps} 
 \fi
 \ifnum\driver=1  #3    \includegraphics{#4.ps}       \fi
 \ifnum\driver=2  #3   \ifnum\mgnf=0
                       \special{#4.ps 1. 1. scale}\fi
                       \ifnum\mgnf=1
                       \special{#4.ps 1.2 1.2 scale}\fi\fi
 \ifnum\driver=5  #3   \fi}
\hfil}}}

\newdimen\xshift \newdimen\xwidth \newdimen\yshift
\def\eqfig#1#2#3#4#5{
\par\xwidth=#1 \xshift=\hsize \advance\xshift 
by-\xwidth \divide\xshift by 2
\yshift=#2 \divide\yshift by 2
\line{\hglue\xshift \vbox to #2{\vfil
\ifnum\driver=0 #3
\special{ps: plotfile #4.ps} 
\fi
\ifnum\driver=1 #3 \includegraphics{#4.ps}\fi
\ifnum\driver=2 #3 \special{
\ifnum\mgnf=0 #4.ps 1. 1. scale \fi
\ifnum\mgnf=1 #4.ps 1.2 1.2 scale\fi}
\fi}\hfill\raise\yshift\hbox{#5}}}

\newskip\ttglue
\def\TIPIO{
\font\setterm=amr7 
\font\settesy=amsy7 \font\settebf=ambx7 
\def \settepunti{\def\rm{\fam0\setterm}
\textfont0=\setterm   
\textfont2=\settesy   
\textfont\bffam=\settebf  \def\bf{\fam\bffam\settebf}
\normalbaselineskip=9pt\normalbaselines\rm
}\let\nota=\settepunti}
\def\annota#1#2{\footnote{${}^#1$}{\nota#2}}

\def\TIPITOT{
\font\twelverm=cmr12
\font\twelvei=cmmi12
\font\twelvesy=cmsy10 scaled\magstep1
\font\twelveex=cmex10 scaled\magstep1
\font\dodici=cmbx10 scaled\magstep1
\font\twelveit=cmti12
\font\twelvett=cmtt12
\font\twelvebf=cmbx12
\font\twelvesl=cmsl12
\font\ninerm=cmr9
\font\ninesy=cmsy9
\font\eightrm=cmr8
\font\eighti=cmmi8
\font\eightsy=cmsy8
\font\eightbf=cmbx8
\font\eighttt=cmtt8
\font\eightsl=cmsl8
\font\eightit=cmti8
\font\sixrm=cmr6
\font\sixbf=cmbx6
\font\sixi=cmmi6
\font\sixsy=cmsy6
\font\twelvetruecmr=cmr10 scaled\magstep1
\font\twelvetruecmsy=cmsy10 scaled\magstep1
\font\tentruecmr=cmr10
\font\tentruecmsy=cmsy10
\font\eighttruecmr=cmr8
\font\eighttruecmsy=cmsy8
\font\seventruecmr=cmr7
\font\seventruecmsy=cmsy7
\font\sixtruecmr=cmr6
\font\sixtruecmsy=cmsy6
\font\fivetruecmr=cmr5
\font\fivetruecmsy=cmsy5
\textfont\truecmr=\tentruecmr
\scriptfont\truecmr=\seventruecmr
\scriptscriptfont\truecmr=\fivetruecmr
\textfont\truecmsy=\tentruecmsy
\scriptfont\truecmsy=\seventruecmsy
\scriptscriptfont\truecmr=\fivetruecmr
\scriptscriptfont\truecmsy=\fivetruecmsy
\def \eightpoint{\def\rm{\fam0\eightrm}
\textfont0=\eightrm \scriptfont0=\sixrm \scriptscriptfont0=\fiverm
\textfont1=\eighti \scriptfont1=\sixi   \scriptscriptfont1=\fivei
\textfont2=\eightsy \scriptfont2=\sixsy   \scriptscriptfont2=\fivesy
\textfont3=\tenex \scriptfont3=\tenex   \scriptscriptfont3=\tenex
\textfont\itfam=\eightit  \def\it{\fam\itfam\eightit}%
\textfont\slfam=\eightsl  \def\sl{\fam\slfam\eightsl}%
\textfont\ttfam=\eighttt  \def\tt{\fam\ttfam\eighttt}%
\textfont\bffam=\eightbf  \scriptfont\bffam=\sixbf
\scriptscriptfont\bffam=\fivebf  \def\bf{\fam\bffam\eightbf}%
\tt \ttglue=.5em plus.25em minus.15em
\setbox\strutbox=\hbox{\vrule height7pt depth2pt width0pt}%
\normalbaselineskip=9pt
\let\sc=\sixrm  \normalbaselines\rm
\textfont\truecmr=\eighttruecmr
\scriptfont\truecmr=\sixtruecmr
\scriptscriptfont\truecmr=\fivetruecmr
\textfont\truecmsy=\eighttruecmsy
\scriptfont\truecmsy=\sixtruecmsy
}\let\nota=\eightpoint}

\newfam\msbfam   
\newfam\truecmr  
\newfam\truecmsy 
\newskip\ttglue
\ifnum\tipi=0\TIPIO \else\ifnum\tipi=1 \TIPI\else \TIPITOT\fi\fi


\let\0=\noindent\let\\=\noindent

\def\ie{\hbox{\it i.e.\ }}\def\eg{\hbox{\it e.g.\ }}
\let\dpr=\partial

\def\*{\vglue0.3truecm}\let\0=\noindent
\let\==\equiv

\def\1{{-1}}
\let\io=\infty

\def\tende#1{\,\vtop{\ialign{##\crcr\rightarrowfill\crcr
              \noalign{\kern-1pt\nointerlineskip}
              \hskip3.pt${\scriptstyle #1}$\hskip3.pt\crcr}}\,}
\def\otto{\,{\kern-1.truept\leftarrow\kern-5.truept\to\kern-1.truept}\,}

\global\newcount\numpunt
\def\XWPR{{\it a priori}}
\def\ap#1{\def\9{#1}{\if\9.\global\numpunt=1\else\if\9,\global\numpunt=2\else
\if\9;\global\numpunt=3\else\if\9:\global\numpunt=4\else
\if\9)\global\numpunt=5\else\if\9!\global\numpunt=6\else
\if\9?\global\numpunt=7\else\global\numpunt=8\fi\fi\fi\fi\fi\fi
\fi}\ifcase\numpunt\or{\XWPR.}\or{\XWPR,}\or
{\XWPR;}\or{\XWPR:}\or{\XWPR)}\or
{\XWPR!}\or{\XWPR?}\or{\XWPR\ \9}\else\fi}


\def\2{{1\over2}}

\def\MM{{\cal M}}
\def\CC{{\cal C}}\def\FF{{\cal F}}\def\HH{{\cal H}}
\def\TT{{\cal T}}\def\NN{{\cal N}}\def\BB{{\cal B}}
\def\VV{{\cal V}}\def\SSS{{\cal S}}
\def\RRR{{\cal R}}\def\LL{{\cal L}}\def\JJ{{\cal J}}

\def\T#1{{#1_{\kern-3pt\lower7pt\hbox{$\widetilde{}$}}\kern3pt}}
\def\VVV#1{{\underline #1}_{\kern-3pt
\lower7pt\hbox{$\widetilde{}$}}\kern3pt\,}
\def\W#1{#1_{\kern-3pt\lower7.5pt\hbox{$\widetilde{}$}}\kern2pt\,}

\def\lis{\overline}

\def\acapo{\hfill\break}


\def\indica{\leaders \hbox to 0.5cm{\hss.\hss}\hfill}

\openout15=\jobname.aux        

\def\guida{\leaders\hbox to 1em{\hss.\hss}\hfill}

\let\h=\eta\let\x=\xi\def\ie{{\it i.e. }}   
\newtoks\footline \footline={\hss\tenrm\folio\hss}
\footline={\rlap{\hbox{\copy200}}\tenrm\hss \number\pageno\hss}
\let\ciao=\bye

\def\annota#1#2{\footnote{${}^#1$}{\nota#2}}

\def\qed{\raise1pt\hbox{\vrule height5pt width5pt depth0pt}}

\def\mm{{\bf m}}
\def\nn{{\bf n}}
\def\aa{{\bf a}}
\def\bb{{\bf b}}

\font\titolo=cmbx12
\font\titolone=cmbx10 scaled\magstep 2
\font\cs=cmcsc10
\font\ss=cmss10
\font\sss=cmss8
\font\crs=cmbx8

\ifnum\mgnf=0
\def\ZZ{\hbox{{\ss Z}\kern-3.3pt{\ss Z}}}
\def\zz{\hbox{{\sss Z}\kern-3.3pt{\sss Z}}}
\fi
\ifnum\mgnf=1
\def\ZZ{\hbox{{\ss Z}\kern-3.6pt{\ss Z}}}
\def\zz{\hbox{{\sss Z}\kern-3.6pt{\sss Z}}}
\fi

\ifnum\mgnf=0
\def\RR{\hbox{I\kern-1.4pt{\ss R}}}
\def\rr{\hbox{{\crs I}\kern-1.3pt{\sss R}}}
\fi
\ifnum\mgnf=1
\def\RR{\hbox{I\kern-1.7pt{\ss R}}}
\def\rr{\hbox{{\crs I}\kern-1.6pt{\sss R}}}
\fi

\ifnum\mgnf=0
\def\openone{\leavevmode\hbox{\ninerm 1\kern-3.3pt\tenrm1}}%
\fi
\ifnum\mgnf=1
\def\openone{\leavevmode\hbox{\ninerm 1\kern-363pt\tenrm1}}%
\fi

\newread\epsffilein    
\newif\ifepsffileok    
\newif\ifepsfbbfound   
\newif\ifepsfverbose   
\newdimen\epsfxsize    
\newdimen\epsfysize    
\newdimen\epsftsize    
\newdimen\epsfrsize    
\newdimen\epsftmp      
\newdimen\pspoints     
\pspoints=1bp          
\epsfxsize=0pt         
\epsfysize=0pt         
\def\epsfbox#1{\global\def\epsfllx{72}\global\def\epsflly{72}%
   \global\def\epsfurx{540}\global\def\epsfury{720}%
   \def\lbracket{[}\def\testit{#1}\ifx\testit\lbracket
   \let\next=\epsfgetlitbb\else\let\next=\epsfnormal\fi\next{#1}}%
\def\epsfgetlitbb#1#2 #3 #4 #5]#6{\epsfgrab #2 #3 #4 #5 .\\%
   \epsfsetgraph{#6}}%
\def\epsfnormal#1{\epsfgetbb{#1}\epsfsetgraph{#1}}%
\def\epsfgetbb#1{%
%
%
\openin\epsffilein=#1
\ifeof\epsffilein\errmessage{I couldn't open #1, will ignore it}\else
%
%
   {\epsffileoktrue \chardef\other=12
    \def\do##1{\catcode`##1=\other}\dospecials \catcode`\ =10
    \loop
       \read\epsffilein to \epsffileline
       \ifeof\epsffilein\epsffileokfalse\else
%
%
          \expandafter\epsfaux\epsffileline:. \\%
       \fi
   \ifepsffileok\repeat
   \ifepsfbbfound\else
    \ifepsfverbose\message{No bounding box comment in #1; using defaults}\fi\fi
   }\closein\epsffilein\fi}%
%
%
\def\epsfclipstring{}
\def\epsfsetgraph#1{%
   \epsfrsize=\epsfury\pspoints
   \advance\epsfrsize by-\epsflly\pspoints
   \epsftsize=\epsfurx\pspoints
   \advance\epsftsize by-\epsfllx\pspoints
%
%
   \epsfxsize\epsfsize\epsftsize\epsfrsize
   \ifnum\epsfxsize=0 \ifnum\epsfysize=0
      \epsfxsize=\epsftsize \epsfysize=\epsfrsize
      \epsfrsize=0pt
%
%
     \else\epsftmp=\epsftsize \divide\epsftmp\epsfrsize
       \epsfxsize=\epsfysize \multiply\epsfxsize\epsftmp
       \multiply\epsftmp\epsfrsize \advance\epsftsize-\epsftmp
       \epsftmp=\epsfysize
       \loop \advance\epsftsize\epsftsize \divide\epsftmp 2
       \ifnum\epsftmp>0
          \ifnum\epsftsize<\epsfrsize\else
             \advance\epsftsize-\epsfrsize \advance\epsfxsize\epsftmp \fi
       \repeat
       \epsfrsize=0pt
     \fi
   \else \ifnum\epsfysize=0
     \epsftmp=\epsfrsize \divide\epsftmp\epsftsize
     \epsfysize=\epsfxsize \multiply\epsfysize\epsftmp   
     \multiply\epsftmp\epsftsize \advance\epsfrsize-\epsftmp
     \epsftmp=\epsfxsize
     \loop \advance\epsfrsize\epsfrsize \divide\epsftmp 2
     \ifnum\epsftmp>0
        \ifnum\epsfrsize<\epsftsize\else
           \advance\epsfrsize-\epsftsize \advance\epsfysize\epsftmp \fi
     \repeat
     \epsfrsize=0pt
    \else
     \epsfrsize=\epsfysize
    \fi
   \fi
%
%
   \ifepsfverbose\message{#1: width=\the\epsfxsize, height=\the\epsfysize}\fi
   \epsftmp=10\epsfxsize \divide\epsftmp\pspoints
   \vbox to\epsfysize{\vfil\hbox to\epsfxsize{%
      \toks0={#1 }%
      \ifnum\epsfrsize=0\relax
        \includegraphics{\the\toks0}%
      \else
        \epsfrsize=10\epsfysize \divide\epsfrsize\pspoints
        \includegraphics{\the\toks0}%
      \fi
      \hfil}}%
\global\epsfxsize=0pt\global\epsfysize=0pt}%
%
%
{\catcode`\%=12 \global\let\epsfpercent=
%
%
\long\def\epsfaux#1#2:#3\\{\ifx#1\epsfpercent
   \def\testit{#2}\ifx\testit\epsfbblit
      \epsfgrab #3 . . . \\%
      \epsffileokfalse
      \global\epsfbbfoundtrue
   \fi\else\ifx#1\par\else\epsffileokfalse\fi\fi}%
%
%
\def\epsfempty{}%
\def\epsfgrab #1 #2 #3 #4 #5\\{%
\global\def\epsfllx{#1}\ifx\epsfllx\epsfempty
      \epsfgrab #2 #3 #4 #5 .\\\else
   \global\def\epsflly{#2}%
   \global\def\epsfurx{#3}\global\def\epsfury{#4}\fi}%
%
%
\def\epsfsize#1#2{\epsfxsize}
%
%
\let\epsffile=\epsfbox

%
%
%
%

\null
\vskip3.truecm
\centerline{\titolone Large deviation rule for Anosov flows}
\vskip1.truecm
\centerline{{\titolo Guido Gentile}$^\dagger$}
\vskip.2truecm
\centerline{$\dagger$ IHES, 35 Route de Chartres, Bures sur Yvette,
France.}
\vskip1.truecm
\line{\vtop{
\line{\hskip1.5truecm\vbox{\advance \hsize by -3.1 truecm
\\{\cs Abstract.}
{\it The volume contraction in dissipative reversible
transitive Anosov flows
obeys a large deviation rule (fluctuation theorem).}} \hfill} }}

\vskip2.truecm

\centerline{\titolo 1. Introduction and formalism}
\*\numsec=1\numfor=1

\\In this paper the results in [G3] are extended to the case of
Anosov flows. The interest and the physical motivation are
explained in [CG1] and [G2]: we briefly review them.

For dissipative systems the existence and properties of
a non equilibrium stationary state are not known in general.
Nevertheless there are systems in which such a state exists
and has been extensively studied: Anosov systems and,
more generally, Axiom $A$ systems.

The content of the {\sl chaotic hypothesis} proposed
in [CG1] generalizing the Ruelle's principle for turbulence, [R5],
is that, as far as only macroscopical quantities have to be computed,
a many particle system in a stationary state out of equilibrium
can be regarded as if it was an Axiom $A$ system.

In [CG1] a theorem of large deviations is heuristically
proven for dissipative reversible transitive Anosov systems,
by using the properties of the stationary state (SRB measure),
and it is shown to
agree with the results of the numerical experiments in [ECM].
A rigorous proof for Anosov diffeomorphisms is performed
in [G3]. As the physical systems which one wants to study
through mathematical models evolve in a continuous way, it can
be interesting to check if the theorem still holds if
one consider Anosov flows instead of diffeomorphisms.
This program is achieved in the present paper: it will be
shown that the study of Anosov flows can be reduced to the study
of Anosov diffeomorphisms (more rigorously of maps
which have all the ``good'' properties of Anosov diffeomorphisms,
in a sense which will be explained below, after Proposition 1.9),
so that the large deviation rule for dissipative reversible
transitive Anosov diffeomorphisms is extended to the case of flows. 

If the systems is Axiom $A$ but not Anosov, something can
still be said: see comments after Theorem 3.6.

\*

In this section we simply review the basic notions and results on
Axiom $A$ flows, Markov partitions and symbolic dynamics,
essentially taken from [B2] and [BR], and
in \S 2 we introduce the SRB measures for Axiom $A$ flows.
Even if in the end we will confine ourselves on Anosov flows,
it can be worthwile to start with more general definitions
(also in view of possible future extensions
to Axiom $A$ flows of the results holding for Anosov flows),
as the discussion in this introductory part
can be carried out with no relevant change
both for Axiom $A$ and Anosov flows.

In \S 3 we consider dissipative reversible Axiom $A$ flows,
study conditions under which they reduce to Anosov flows,
and state the fundamental result of the paper (a large deviation rule
for the volume contraction in dissipative reversible
transitive Anosov flows), which will be proven in \S 4.

\*

Let $M$ be a differentiable ($C^{\io}$) compact Riemannian manifold
and $f^t\!\!: M \to M$ a differentiable flow.

\*

\\{\bf 1.1.} {\cs Definition.} {\it A closed $f^t$-invariant
set $X\subset M$ containing no fixed points is {\sl hyperbolic}
if the tangent bundle restricted to $X$ can be written as the
Whitney sum\annota{1}{\nota
That is for each $x\in X$, the decomposition \equ(1.1) becomes
$T_xM=E_x\oplus E_x^s \oplus E_x^u$.}
of three $Tf^t$-invariant continuous subbundles
$$ T_X M = E + E^s + E^u \; , \Eq(1.1) $$
where $E$ is the one-dimensional bundle tangent to the flow, and
$$ \eqalign{
(a) \kern.5truecm &
\| T f^t w \| \le c\,e^{-\l t}\|w\| \; , \quad \hbox{for } w\in E^s \; ,
\quad t\ge 0 \; , \cr
(b) \kern.5truecm &
\| T f^{-t} w \| \le c\,e^{-\l t}\|w\| \; , \quad \hbox{for } w\in E^u \; ,
\quad t\ge 0 \; , \cr} \Eq(1.2) $$
for some positive constants $c,\l$; $\|\cdot\|$ denotes the norm
induced by the Riemann metric.}

\*

More generally, if $Y$ is the union of a hyperbolic set
as above and a finite number of hyperbolic fixed
points, we also say that $Y$ is hyperbolic, [ER], \S F.2.


\*

\\{\bf 1.2.} {\cs Definition.} {\it
A closed $f^t$-invariant set $\L$ is a {\sl basic hyperbolic set} if
\acapo
(1) $\L$ contains no fixed points and is hyperbolic;
\acapo
(2) the periodic orbits of $f^t|\L$ are dense in $\L$;
\acapo
(3) $f^t|\L$ is topologically transitive;\annota{2}{\nota
A flow $f^t\!:\L \to \L$ is topologically transitive if,
for all $U$, $V\subset \L$ open nonempty,
$U \cap f^t V \neq \emptyset$ for some $t>0$.}
\acapo
(4) there is an open set $U \supset \L$ with $\L = \cap_{t\in{\rr}}
f^tU$.}

\*

Definition 1.2 is taken from [BR], \S 1. Usually one defines
a basic hyperbolic set as a set which either
satisfies Definition 1.2 or is a hyperbolic fixed point.
In the following we will be interested in basic
hyperbolic sets which are not a single point:
this motivates Definition 1.2.

\*

\\{\bf 1.3.} {\cs Definition.} {\it
A basic hyperbolic set $\L$ for which the set $U$ in item (4) can be
chosen satisfying $f^tU\subset U$ for all $t\ge t_0$, for fixed $t_0$,
is defined to be an {\sl attractor}.}

\*

A point $x\in M$ is called {\sl nonwandering} if,
for every neighbourhood $V$ of $x$ and every $t_0\in\RR$,
there is a $t>t_0$ such that
$$ f^tV \cap V \neq \emptyset \; . $$
The {\sl nonwandering set} is defined as
$$ \O = \{ x \in M : \hbox{$x$ is nonwandering} \} \; . $$

\*

\\{\bf 1.4.} {\cs Definition.} {\it
A flow $f^t\!\!:M\to M$ is said to satisfy {\sl Axiom $A$}
if the nonwandering set $\O$ is the disjoint union of
a set satisfying (1) and (2) of Definition 1.2
and a finite number of hyperbolic fixed points.}

\*

Smale's {\sl spectral decomposition theorem} ([Sm],
Theorem 5.2; see also [PS], Theorem 2.1)
states that, if the flow $f^t\!\!:M\to M$ satisfies
Axiom $A$, and if we denote by $\FF$ the
set of hyperbolic fixed points in $\O$,
then $\O\setminus\FF$ is the disjoint union of
a finite number of basic hyperbolic sets.

\*

\\{\bf 1.5.} {\cs Definition.} {\it
A flow $f^t\!\!:M\to M$ is an {\sl Anosov flow} if $M$ is hyperbolic.}

\*

An Anosov flow satisfies Axiom A
(Anosov's {\sl closing lemma}, [A]; see also [B3], \S 3.8).
By Smale's spectral decomposition theorem,
given an Anosov flow $f^t\!\!:M\to M$,
then one can decompose $\O=\cup_{j=1}^N \L_j$,
where $\L_1,\ldots,\L_N$ are basic hyperbolic sets,
and one can consider the restriction $f^t|\L_j$, $\forall j=1,\ldots,N$,
which is topologically transitive.
If one has $\O=M$, then each $f^t|\L_j$ is a
{\sl transitive Anosov flow}.\annota{3}{\nota 
Note that there are Anosov flows for which $\O\neq M$, [FW]:
such flows are obviously non transitive.
In the case of maps, the identity $\O=M$ is
conjectured to hold for all Anosov diffeomorphisms, [Sm],
Problem 3.4.}

Standard examples of Axiom $A$ flows are the suspension of an
Axiom $A$ diffeomorphism, \eg the solenoid, [Sm],
and the geodesic flow on a compact
manifold with negative curvature, [A], which is an Anosov flow,
(see also [B4]).

\*

Let $\L$ be a basic hyperbolic set.
For any $x\in\L$, the stable and unstable manifolds are defined as
$$ \eqalign{
W_x^s & = \{ y\in M \; : \; \lim_{t\to\io} d(f^tx,f^ty) = 0 \} \; , \cr
W_x^u & = \{ y\in M \; : \; \lim_{t\to\io} d(f^{-t}x,f^{-t}y) = 0 \}
\; , \cr} \Eq(1.3) $$
where $d$ is the distance induced by the Riemann metric, and
$$ W_\L^s = \bigcup_{x\in\L}W_x^s \; , \quad
W_\L^u = \bigcup_{x\in\L}W_x^u \; . \Eq(1.4) $$
If $\L$ is an attractor, $W_\L^s$ is its {\sl basin}: $\lim_{t\to\io}
d(f^tx,\L) = 0$ $\forall x\in W_{\L}^s$.


%
%

\*

For $x\in\L$, we set
$$ \eqalign{
W_{x,\e}^s & = \{ y \in W_x^s \; : \; d(f^tx,f^ty) \le \e
\; \forall t\ge 0 \} \; , \cr
W_{x,\e}^u & = \{ y \in W_x^u \; : \; d(f^{-t}x,f^{-t}y) \le \e
\; \forall t\ge 0 \} \; . \cr} $$

\*

Let $D$ be a differentiable closed disk (\ie a closed
element of a $C^{\io}$ manifold of dimension $\hbox{dim}(M)-1$, if
$\hbox{dim}(M)$ is the dimension of $M$),
containing a point $x\in\L$
and trasverse to the flow. For any closed subset $T\subset D$
containing $x$ and for any $y\in T$ such that
$d(x,y) \le \a_1$ for a suitable $\a_1$, let us define
$\langle x,y \rangle$ as the intersection
$W^s_{f^vx,\e}\cap W^u_{y,\e}$, for a suitable
$v$, (by choosing $\a_1$ small enough, one can always
take $|v|\le \e$): such an intersection is well defined
(\ie it is a single point) and lies in $\L$, [Sm].
Then let us introduce the {\sl canonical coordinate}
$\langle x,y \rangle_D$ as the {\sl projection} of
$\langle x,y \rangle$ on $D$, [B2], \S 1:
this means that there exists a constant $\x\ge0$ such that,
for $|r|\le \x$, $f^r\langle x, y \rangle_D$
$=$ $\langle x,y \rangle$.
The subset $T$ is called a {\sl rectangle} if
$\langle x,y \rangle_D \in T$ for any $x,y\in T$,
and in this case we can define
$\langle x,y \rangle_T\=\langle x,y \rangle_D$.

%
\midinsert
$$ \vcenter{\epsfysize=6.cm \epsffile{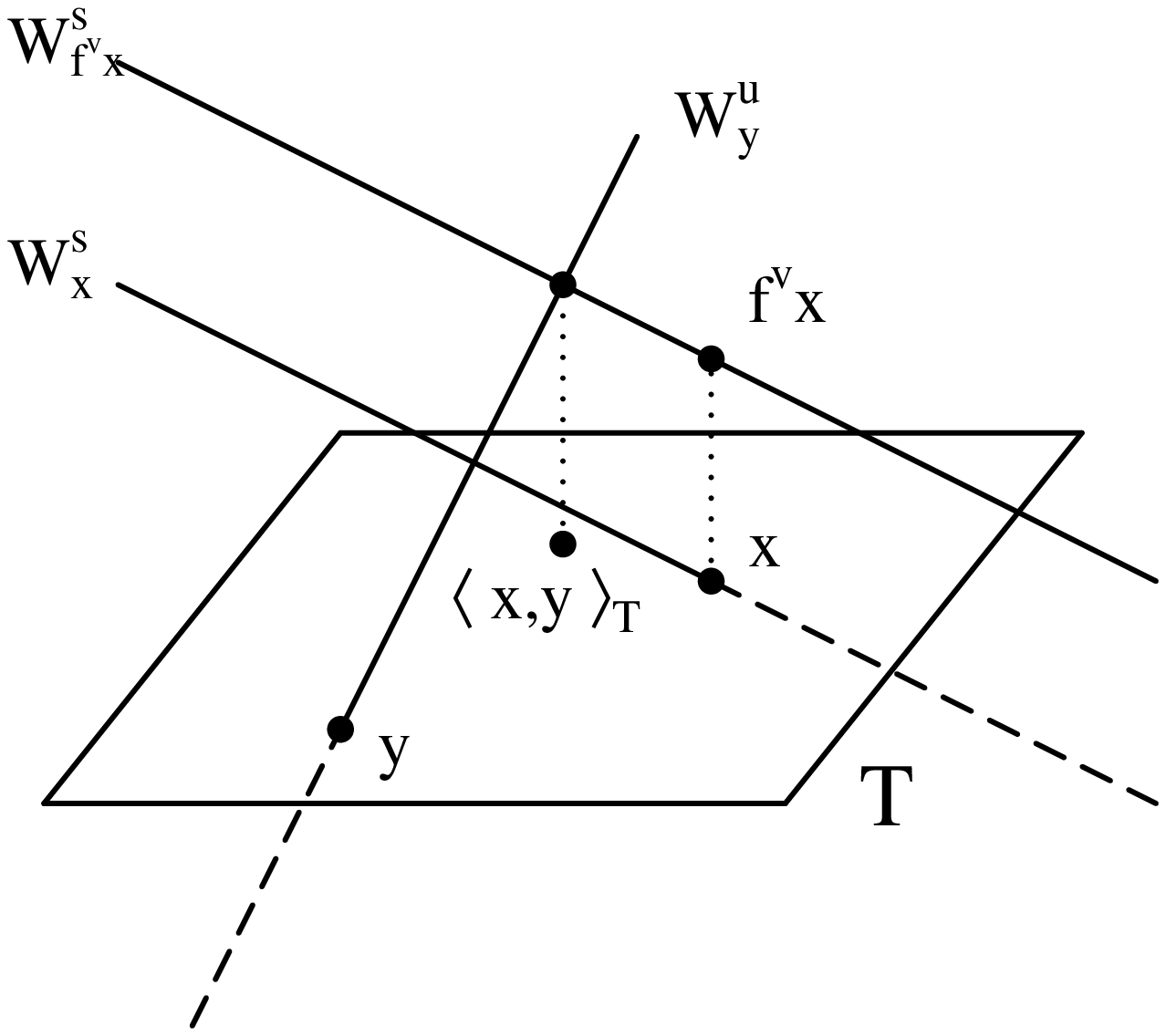} } $$
\centerline{{\cs Fig.1.} {\it Canonical coordinates.}}
\endinsert
%

Then, for $x\in T$, we set
$W^s_x(T)=\{\langle x,y\rangle \; : \; y \in T \}$ and
$W^u_x(T)=\{\langle y,x\rangle \; : \; y \in T \}$: we can say
that $W^s_x(T)$ and $W_x^u(T)$ are the projection of the
stable manifold and, respectively, of the unstable manifold
of $x$ on the rectangle $T$ containing $x$ (the
projection is meant along the flow), and $T$ becomes the
direct product of $W_x^s(T)$ and $W_x^u(T)$.

\*

\\{\bf 1.6.} {\cs Definition.} {\it
Choose a basic hyperbolic set $\L$.
We call $\TT=\{T_1,\ldots,T_\NN \}$ a {\sl proper family of
rectangles}, if there are positive constants $\a$ and $\a_1$ such that
\acapo
(1) $T_j\subset\L$ is a closed rectangle;
\acapo
(2) if $\G(\TT)=\bigcup_{j=1}^{\NN} T_j$, then
$\L \subset \bigcup_{0\le t\le\a}f^{-t}\G(\TT)$;
\acapo
(3)} $T_j\subset \hbox{int}\,D_j$, {\it where $D_j$ is a $C^{\io}$
closed disk transverse to the flow, such that: (3.1)}
$\hbox{diam}\,(D_j)\le \a_1$ {\it $\forall j$, (3.2)}
$\hbox{dim}\,(D_j) = \hbox{dim}\,(M)-1$ {\it $\forall j$,
(3.3)} $T_j =
\overline{\hbox{int}\,T_j}$
{\it $\forall j=1,\ldots,\NN$, (where} $\hbox{int}\,T_j$
{\it is the interior of $T_j$ as a subset of $\L\cap D_j$),
and (3.4) for $i\neq j$, at least one of the sets
$D_i\cap \bigcup_{o\le t\le\a}f^tD_j $ and
$D_j\cap \bigcup_{o\le t\le\a}f^tD_i$ is empty (in particular
$D_i\cap D_j=\emptyset$).}

\*

The above definition is taken from [B2], Definition 2.1.
[In [B2], $\a_1=\a$ and $\TT$ is called a proper family
of rectangles ``of size $\a$''; in general it can be useful
to consider $\a$ and $\a_1$ as independent parameters, so that
one can change one of them without affecting the other one.]

Note that, if the flow $f^t\!\!:M\to M$ is a topologically mixing
Anosov flow,\annota{4}{\nota
A transitive Anosov flow is said to be topologically mixing if
the stable and unstable manifolds $W_x^s$ and $W_x^u$ are dense in $M$
for some (and then for each) $x\in M$.}
then $E^s$ and $E^u$ are not jointly integrable
(\ie $E^s\oplus E^u$ is not integrable, [Pl], Proposition 1.6),
[Pl], Lemma 1.4, Lemma 1.5, Theorem 1.8.
This means that, if $\e'$ is so chosen that for any $x\in\L$,
$y\in W^u_{x,\e}$ and $\x\in W^s_{x,\e'}$ one has
$W^u_{\x,\e}\cap\bigcup_{-\e\le t\le\e}f^t W^s_{y,\e}\neq\emptyset$,
then $W^u_{\x,\e}\cap W^s_{y,\e}=\emptyset$.
Therefore, in such a case, the disks
$D_j$'s can {\it not} be constructed so that
the conditions $W_{x,\e}^s(T)=W_{x,\e}^s\cap T$ and
$W_{x,\e}^u(T)=W_{x,\e}^u\cap T$ are simultaneously possible.

Given $x\in\G(\TT)$, let $t'(x)$ be the first positive
time required for $f^tx$ to cross $\G(\TT)$.
If $x\in T_i$, for some $i=1,\ldots,\NN$,
then $f^{t'(x)}x\in T_j$, for some $j\neq i$;
set $t(x)=t'(x)$ for $x\in \hbox{int}\,T_i$ and
extend it by continuity to the boundaries of $T_i$.
Then define
$\HH_\TT x = f^{t(x)}x$, for any $x\in\G(\TT)$, [B2], \S 2: $t(x)$
is called {\sl ceiling function} and $\HH_\TT$ {\sl Poincar\'e map}.
There exists a $t_0\in(0,\a)$ such that $t(x)>t_0$ $\forall x\in \G(\TT)$.

The function $\HH_{\TT}$ is continuous on
$$ \G'(\TT) = \{ x \in \G(\TT) \; : \; \HH_{\TT}^kx \in
\bigcup_{j=1}^{\NN} \hbox{int}\,T_j \;
\forall k \in \ZZ \} \; , \Eq(1.5) $$
and $\G'(\TT)$ is dense in $\G(\TT)$, being a countable
intersection of dense open subsets ({\sl Baire's theorem},
[Bb], Ch. IX, \S 5.3).

\*

\\{\bf 1.7.} {\cs Definition.} {\it
A proper family of rectangles $\TT$ is called a 
{\sl Markov partition} (or {\sl Markov family},
or {\sl Markov pavement}), if
\acapo
(1) for $x\in T_i$, $\HH_\TT x \in T_j$,
one has $\HH_\TT y \in T_j$ $\forall y \in W^s_x(T_i)$;
\acapo
(2) for $x\in T_i$, $\HH_{\TT}^{-1} x \in T_j$,
one has $\HH_{\TT}^{-1} y \in T_j$ $\forall y \in W^u_x(T_i)$.}

\*

The above definition is taken from [B2], Definition 2.3.

Define
$$ \eqalign{ & \dpr^s\TT = \bigcup_{j=1}^{\NN} \dpr^s T_j \; ,
\quad \dpr^u\TT = \bigcup_{j=1}^{\NN} \dpr^u T_j \; ,
\quad \dpr\TT = \dpr^s\TT \cup \dpr^u\TT \; , \cr
& \D^s\TT = \bigcup_{0\le t\le\a} f^t \dpr^s\TT \; , \quad
\D^u\TT = \bigcup_{0\le t\le\a} f^{-t} \dpr^u\TT \; ; \cr} $$
where
$$ \eqalign{
\dpr^s T & = \{ \langle y_1, y_2 \rangle_T \; : \;
y_1\in \dpr W_{x,\e}^u(T) \hbox{ and } y_2\in W_{x,\e}^s(T) \}\; , \cr
\dpr^u T & = \{ \langle y_1, y_2 \rangle_T \; : \;
y_1\in W_{x,\e}^u(T)\hbox{ and } y_2\in\dpr W_{x,\e}^s(T) \}\;,\cr} $$
if $\dpr W_{x,\e}^u(T)$ and $\dpr W_{x,\e}^s(T)$ denote the boundaries
of $W_{x,\e}^u(T)$ and $W_{x,\e}^s(T)$ as subsets,
respectively, of $W_{x}^u(T)\cap\L$ and $W_{x}^s(T)\cap\L$.

\*

\\{\bf 1.8.} {\cs Proposition.}
{\it If $\TT$ is a Markov partition, one has
$f^t\D^s\TT \subset\D^s\TT$ and $f^{-t}\D^u\TT \subset \D^u\TT$
$\forall t\ge0$.}

\*

The proof is in [B2], Proposition 2.6.

Then the following fundamental result is proven in [B2], Theorem 2.5.

\*

\\{\bf 1.9.} {\cs Proposition.}
{\it Any basic hyperbolic set $\L$ admits a
proper family of rectangles
which is a Markov partition.}

\*

By construction, the discontinuity set of $\HH_\TT$, \ie
$\G(\TT)\setminus\G'(\TT)$, is covered by the evolution of some stable
and unstable manifolds, so that $\HH_{\TT}\!\!:\G'(\TT)\to\G'(\TT)$ can
be studied as it was an Axiom $A$ diffeomorphism, (see [B3] for a review).

\*

\\{\bf 1.10.} {\it Symbolic dynamics.}
Let us introduce a $\NN\times\NN$ matrix $A$ such that
$$ A_{ij} = \cases{ 1 & if there exists $x\in \hbox{int}\,T_i$ such that
$\HH_\TT x \in \hbox{int}\,T_j$, \cr
0 & otherwise, \cr} $$
({\sl transition matrix}),
and let us define the {\sl space of the compatible strings}
$$ \MM = \{ \mm \= \{m_i\}_{i\in{\zz}} \; : \; m_i\in\{1,\ldots,\NN\}
\, , \; A_{m_im_{i+1}}=1 \; \forall i\in{\ZZ} \} , \Eq(1.6) $$
and the map $\s \! : \MM \to \MM$ by $\s\mm = \{m_i'\}_{i\in{\zz}}$,
where $m_i'=m_{i+1}$. If $\{1,\ldots,\NN\}$ is given the
discrete topology and $\{1,\ldots,\NN\}^{\zz}$ the product
topology, $\MM$ becomes a compact metrizable space and $\s$ a
topologically transitive homeomorphism ({\sl subshift of
finite type}); furthermore, because of the transitivity of $f^t$,
$\s$ can be supposed to be topologically mixing,\annota{5}{\nota
A homeomorphism $f\! : X\to X$ is topologically mixing if, for
all $U$, $V \subset X$ open nonempty, $U\cap f^n V \neq \emptyset$ for
all sufficiently large $n$.} [BR], Lemma 2.1.
A metric on $\MM$ can be $d(\mm,\nn)=d_1e^{-d_2N}$, with $d_1,d_2>0$,
if $m_i=n_i$, $\forall |i|\le N$, [B2], \S 1.

For $\psi \!:  \MM \to {\RR}$ a positive continuous function,
\ie $\psi\in C(\MM)$, and 
$$ Y = \{(\mm,s)\; :\; s\in[0,\psi(\mm)],\;\mm\in\MM\} \; , \Eq(1.7) $$
identify the points $(\mm,\psi(\mm))$ and $(\s\mm,0)$ for all
$\mm\in\MM$, so obtaining a new compact metric space $\L(A,\psi)$, [BW].
If $q: \; Y \to \L(A,\psi)$ is the quotient map, then the
{\sl suspension flow} (or {\sl special flow}) $g^t\! :\L(A,\psi) \to
\L(A,\psi)$ is defined as
$$ g^tq(\mm,s) = q(\s^k\mm,v) \; , $$
where $k$ is chosen so that
$$ v = t + s - \sum_{j=0}^{k-1} \psi(\s^j\mm) \; \in [0,\psi(\s^k\mm)]
\; . $$

%
\midinsert
$$ \vcenter{\epsfysize=6.cm \epsffile{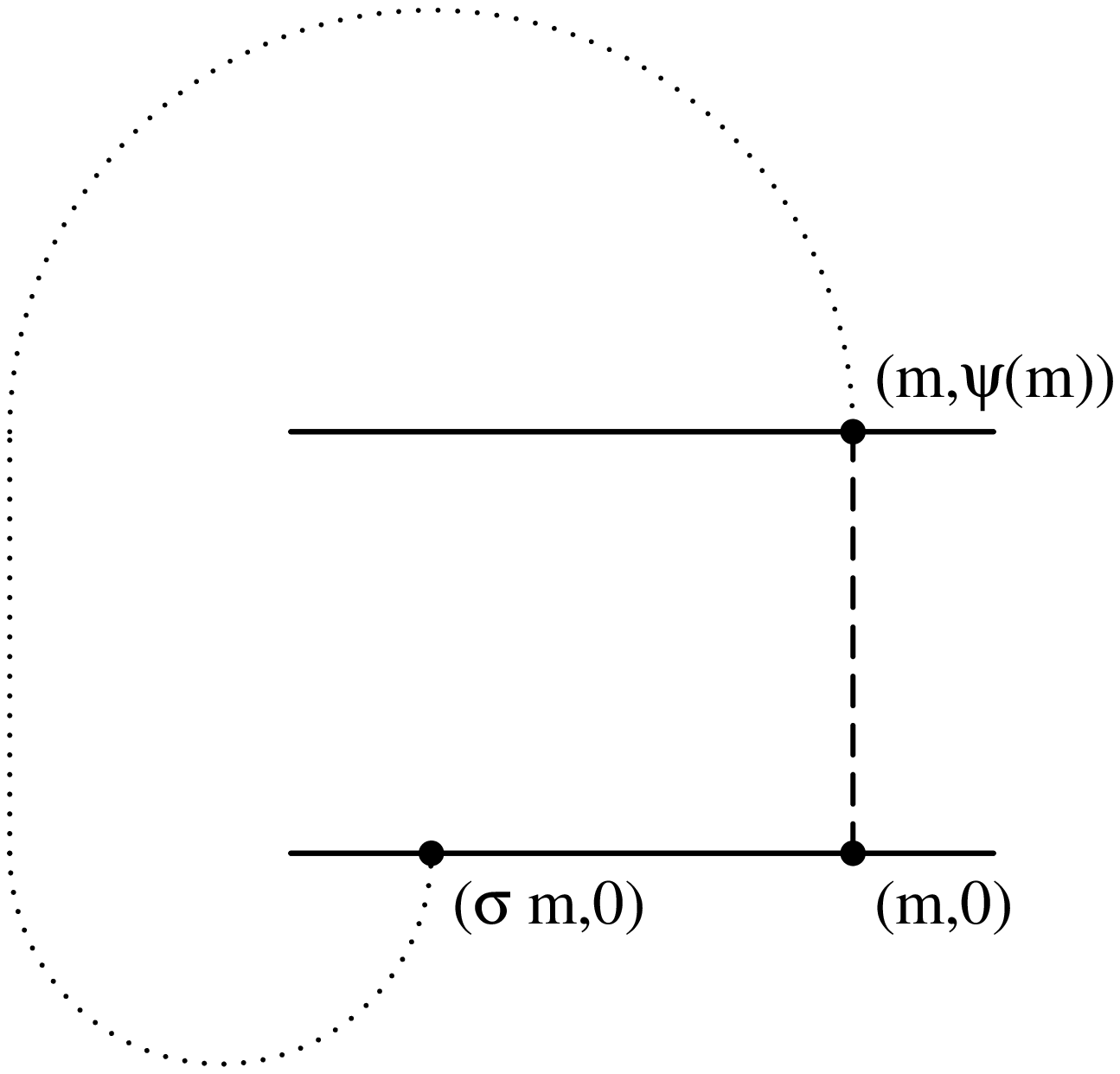} } $$
\centerline{{\cs Fig.2.} {\it Suspension flow.}}
\endinsert
%

For $\psi\in C(\MM)$, let
$$ \hbox{var}_N \psi = \sup \{ |\psi(\mm) - \psi(\nn)|\;:\;
\mm,\nn\in\MM , \; m_i=n_i \; \forall |i|\le N \} \; $$
and let
$$ \FF = \{ \psi \in C(\MM)\; : \; \exists \; c_1,c_2>0
\hbox{ so that var}_N\psi\le c_1\,e^{-c_2N}
\; \forall N\ge0\}. \Eq(1.8) $$

A flow $g^t$ with $\psi\in\FF$ is called a {\sl hyperbolic symbolic
flow}, [B2], Definition 1.3, and its class is the same as the class
of all one-dimensional basic sets for flows, [B1].

\*

\\{\bf 1.11.} {\cs Proposition.} {\it
If $\L$ is a basic hyperbolic set, there exists a positive
$\psi\in\FF$ and a continuous surjection $\r\!:\;\L(A,\psi)\to\L$
({\sl symbolic code})
such that $\r \circ g^t = f^t \circ \r$.}

\*

The proof is in [B1], \S 2.
If $N=\bigcup_{t=-\io}^{\io} f^t\dpr\TT$, and we set
$\L_0=\L\setminus N$ and $\MM_0=\L(A,\psi)\setminus\r^{-1}(N)$, then
$\r$ is a continuous bijection between $\L_0$ and $\MM_0$.
Note that, if $x\in\G(\TT)$, then $x=\r(\mm,0)$,
$\mm\in\MM$, and one has $t(x)=\psi(\mm)$.

\vskip1.truecm

\centerline{\titolo 2. Equilibrium states and SRB measures}
\*\numsec=2\numfor=1

\\Given a homeomorphism $f$, $M(f)$ denotes the set of $f$-invariant
Borel probability measures; if $F=\{f^t\}_{t\in{\rr}}$, then
$M(F)=\bigcap_{t\in{\rr}}M(f^t)$. If $g^t$ is the suspension flow,
we set $G=\{g^t\}$.

Let $\L$ be a basic hyperbolic set.
For $x\in\L$, if $E_x^s$ and $E_x^u$ denote, respectively,
the subbundles tangent to $W_x^s$ and $W_x^u$ in $x$,
let $\l_{0,t}(x)$, $\l_{u,t}(x)$ and $\l_{s,t}(x)$
be the jacobians of the linear maps, respectively,
$Df^t\!: E_x\to E_{f^tx}$,
$Df^t\!: E_x^u\to E_{f^tx}^u$
and $Df^t\!: E_x^s\to E_{f^tx}^s$, and
$\l_t(x)$ $=$ $\l_{0,t}(x)$ $\l_{s,t}(x)$ $\l_{u,t}(x)$
$\chi_{t}^1(x)$ $\chi_{t}^2(x)$,
where $\chi_t^1(x)=
\sin[\psi^1(f^tx)]/\sin[\psi^1(x)]$ and
$\chi_t^2(x)=\sin[\psi^2(f^tx)]/\sin[\psi^2(x)]$,
being $\psi^1(x)$ the angle between $E^s_x$ and $E^u_x$ in $x$,
and $\psi^2(x)$ the angle between
$E^s_x \oplus E^u_x$ and the flow direction in $x$.

Note that $\l_{u,t+t'}(x)=\l_{u,t}(f^{t'}x)\,\l_{u,t'}(x)$ and analogous
relations hold for $\l_{s,t}$, $\l_{0,t}$, $\chi_t^1$ and $\chi_t^2$:
so that all such quantities are {\sl cocycles},
in the sense of [R4], Definition B2.

By the transversality properties and the absence of fixed points
of the basic hyperbolic sets (see Definitions 1.1 and 1.2),
there exists a positive constant $B_1$ such that
$B^{-1}_1<\chi_t^1(x),\chi_t^2(x),\l_{0,t}(x)<B_1$,
for each $t\in\RR$, $x\in \L$.

Then we define
$$ \eqalign{
\f^{(u)}(x) & = - {d\over dt} \ln \l_{u,t}(x)\Big|_{t=0} \; , \cr
\f^{(s)}(x) & = {d\over dt} \ln \l_{s,t}(x) \Big|_{t=0} \; , \cr}
\Eq(2.1) $$
so that, for $x=\r(\mm,0)$ and $\t_k(x) = \sum_{j=0}^{k-1} t(\r(\s^j\mm,0))
\= \sum_{j=0}^{k-1} \psi(\s^{j}\mm)$, if
$$ \eqalign{
J_{u,k}(x) & =\l_{u,\t_k(x)}(x) \; , \quad 
J_{s,k}(x) =\l_{s,\t_k(x)}(x) \; , \quad 
J_k(x)(x) = \l_{\t_k(x)}(x) \; , \cr
J_u(x) \= & J_{u,1}(x) \; , \quad\;
J_s(x) \= J_{s,1}(x) \; , \quad\;
J(x) \= J_{1}(x) \; , \cr} $$
we obtain
$$ \eqalign{
\ln J_{u,k}(x) & = - \int_0^{\t_k(x)} dt \, \f^{(u)}(f^tx) 
= \sum_{j=0}^{k-1} \ln J_u(\HH^j_\TT x) \; ,
\quad \ln J_u(x) = \int_0^{\t_1(x)}dt\,\f^{(u)}(f^tx) \; , \cr
\ln J_{s,k}(x) & = - \int_0^{\t_k(x)} dt \, \f^{(s)}(f^tx) 
= \sum_{j=0}^{k-1} \ln J_s(\HH^j_\TT x) \; ,
\quad \ln J_s(x) = \int_0^{\t_1(x)}dt\,\f^{(s)}(f^tx) \; , \cr} $$
and we can define for even $k$
$$ \eqalign{
\ln \JJ_{u,k}^{-1} & \= \int_{\t_{-k/2}(x)}^{\t_{k/2}(x)} dt \,
\f^{(u)}(f^tx) = \sum_{j=-k/2}^{k/2-1} \ln J_u^{-1}(\HH_\TT^jx) \; , \cr
\ln \JJ_{s,k} & \= - \int_{\t_{-k/2}(x)}^{\t_{k/2}(x)} dt \,
\f^{(s)}(f^tx) = \sum_{j=-k/2}^{k/2-1} \ln J_s(\HH_\TT^jx) \; . \cr}
\Eq(2.2) $$

\*

For $\n\in M(\s)$ and $m$ Lebesgue measure,
$\m_\n \= [\n\times m\,(Y)]^{-1}\,\n\times m|Y$ gives a
probability measure on $\L(A,\psi)$, as $\n \times m$
gives measure zero to the identifications on $Y\to \L(A,\psi)$,
so that no ambiguity can arise. [$Y$ is defined in \equ(1.7).]

\*

\\{\bf 2.1.} {\cs Proposition.} {\it
There exists a measure $\m_+\in M(G)$, which is the
unique equilibrium state for $\f^{(u)}$ with respect to
$G$, ergodic and positive on nonempty open sets,
({\sl forward SRB measure}), and a measure $\m_-\in M(G)$, which is the
unique equilibrium state for $\f^{(s)}$ with respect to
$G^{-1}$, ergodic and positive on nonempty open sets,
({\sl backward SRB measure}). One has $\m_\pm=\m_{\n_\pm}$,
$\n_{\pm}\in M(\s)$, where $\n_+$ and $\n_-$ are,
respectively, the unique equilibrium states for
$\F^{(u)}$ and $\F^{(s)}$, if
$$ \F^{(u)}(\mm) = \int^{\psi(\mm)}_0 dt \, \f^{(u)}(\r(\mm,t)) \;
, \qquad \F^{(s)}(\mm) = \int^{\psi(\mm)}_0 dt \, \f^{(s)}(\r(\mm,t))
\; , $$
with respect to $\s$.}

\*

The proof for $\m_+$ is in [BR], Proposition 3.1,
(where the definition of equilibrium state can also be found),
and for $\m_-$ can be carried out in the same way,
by studying $f^{-t}$ instead of $f^t$ and taking into account
the fact that the unstable manifolds for $f^t$ becomes the stable
manifolds for the opposite flow $f^{-t}$ and {\it viceversa}.

For transitive Anosov flows, we have the following result,
(note that if $f^t\!\!:M\to M$ is a transitive Anosov flow, then
$M=\L$, see comments after Definition 1.5).

\*

\\{\bf 2.2.} {\cs Proposition.} {\it
Let $f^t\!\!:M\to M$ be a transitive Anosov flow .
The volume measure $\m_0$ on $\L$ admits the representation
$\m_0 = \m_{\n_0}$, with $\n_0$
formally proportional to $\exp[-H(\mm)]$, where the formal Hamiltonian
$H(\mm)$ is given by
$$ H(\mm)=\sum_{j=-\io}^{-1} h_-(\s^j\mm)+h_0(\mm)+\sum_{j=0}^\io
h_+(\s^k \mm) \; , \Eq(2.3) $$
with
$$ \cases{
h_-(\mm) = -\ln J_s(\r(\mm,0)) \; , \quad
h_+(\mm)=\ln J_u(\r(\mm,0)) \; , & \cr
h_0(\mm) = -\ln \chi (\r(\mm,0)) \; , & \cr} $$
%
being $\chi(\r(\mm,0))$
bounded between two constants, which we can take as $B_2^{-1}<B_2$.
If $\n_+$, $\n_-$ are the Gibbs states with formal Hamiltonians
$$ H_+(\mm)=\sum_{j=-\io}^\io h_+(\s^j\mm)\;,\qquad
H_-(\mm)=\sum_{j=-\io}^\io h_-(\s^j\mm) \; , \Eq(2.4) $$
then $\m_{\n_+}$, $\m_{\n_-}$ are, respectively,
the forward and backward SRB measures $\m_+$, $\m_-$ on $\L$.}

\*

A statement similar to Proposition 2.2 holds for
diffeomorphisms (see [G3]), and follows from
the analysis in [Si1,Si2,Si3] and [G2,G3].
In Appendix A1, we show how to reduce the discussion of the flows
to the case of diffeomorphisms, so that Proposition 2.2 follows.

\*

\\{\bf 2.3.} {\it Remark.} Note that, unlike the SRB measures,
the volume measure is not translation invariant
(so that it is not really a Gibbs state, see [R3]):
the non translation invariance is due not to any
symmetry breaking phenomenon, but simply to the
fact that the potential ``to the right'' is different from the
potential ``to the left''.

\*

If $g\!:M\to {\RR}$ is smooth,
the function $g(\r(\mm,0))$, $\mm\in\MM$,
can be represented in terms of suitable
functions $\g_k(m_{-k},\ldots,m_k)$ as
$$ g(\r(\mm,0)) = \sum_{k=1}^\io \g_k(m_{-k},\ldots,m_k)\;,\qquad
|\g_k(m_{-k},\ldots,m_k)|\le \G\, e^{-\l k} \; , $$
where $\G>0,\l>0$.
In particular $h_\pm$ (and $h_0$) enjoy the
above property ({\sl short range}), by the properties of the Markov
partition introduced in \S 1.

\*

\\{\bf 2.4.} {\cs Proposition.} {\it For any smooth function}
$g\!:M\to\RR$, {\it if $\L$ is an attractor for the Axiom $A$ flow
$f^t\!:M\to M$, and $W_\L^s$ is its basin, one has
$$ \lim_{T\to\io} {1\over T} \int_0^T dt \, g(f^{t}x) = \int_{\L}
\m_{+} (dy) \, g(y) $$
for $\m_0$-almost all $x\in W_\L^s$. Analogously, if $\L'$ is an
attractor for the opposite flow $f^{-t}\!\!:M\to M$, one has
$$ \lim_{T\to\io} {1\over T} \int_0^T dt \, g(f^{-t}x) = \int_{\L'}
\m_{-} (dy) \, g(y) $$
for $\m_0$-almost all $x\in W_{\L'}^u$, where $W_{\L'}^u$ is the
basin of $\L'$: $\lim_{t\to\io}d\,(f^{-t}x,\L')=0$ $\forall x\in
W_{\L'}^u$.}

\*

The proof is in [BR], Theorem 5.1.

\*

Consider transitive Anosov flows.

Let us construct the Markov partition $\TT_L = \bigvee_{j=-L}^{L}
\HH_{\TT}^{-j} \TT$, (this means that, if
$\mm_{[-L,L]}$ $\=$ $(m_{-L},\ldots,m_L)$ and $T_{\mm_{[-L,L]}}$ $=$
$\bigcap_{j=-L}^{L} \HH_{\TT}^{-j} T_{m_j}$, for $T_{m_j}\in\TT$
$\forall j$ $=$ $-L,\ldots,L$, then $T_{\mm_{[-L,L]}}\in\TT_L$),
and let $x_{\mm_{[-L,L]}^0}$
be a suitable point in $T_{\mm_{[-L,L]}}$,
where $\mm_{[-L,L]}^0\in\MM$, (\ie $\mm_{[-L,L]}^0$ is
a compatible string), with $(\mm_{[-L,L]}^0)_i$ $=$
$(\mm_{[-L,L]})_i$, $\forall |i|\le L$; for instance we can choose
the symbols corresponding to the sites $|j|\ge\pm(L+1)$ such that
$A_{m_jm_{j+1}}=1$, so that the dependence on $\mm_{[-L,L]}$ is
only via the symbols $m_{\pm L}$.

We can define
$$ \int_{\L}
\m_{L,k}(dx) \, g(x) = {\sum_{\mm_{[-L,L]}} \JJ_{u,k}^{-1}
(x_{\mm_{[-L,L]}^0}) \, \int_0^{\psi(\mm_{[-L,L]}^0)} dt \,
g(f^tx_{\mm_{[-L,L]}^0}) \over
\sum_{\mm_{[-L,L]}} \JJ_{u,k}^{-1}
(x_{\mm_{[-L,L]}^0}) \, \psi(\mm_{[-L,L]}^0) } \, \; , \Eq(2.5) $$
which is called {\sl approximating distribution for $\m_+$.} In fact
the following result holds.

\*

\\{\bf 2.5.} {\cs Proposition (Approximation theorem).} {\it
Let $f^t\!\!:M\to M$ be a transitive Anosov flow.
If $\m_{L,k}$ is defined as in \equ(2.5), then, for any smooth function}
$g\!: M \to \RR$, {\it one has
$$ \lim_{k\to\io \atop L\ge k/2} \int \m_{L,k}(dx) \, g(x) =
\int \m_+(dx) \, g(x) \; , $$
where $\m_+$ is the forward SRB measure.}

\*

The measure $\m_{L,k}$ can be written as
$\m_{L,k}\=\m_{\n_{L,k}}$, where $\n_{L,k}$ is the approximating
distribution for $\n_+ \in M(\s)$ defined on $\MM$.
In the following we shall use the notation
$$ \int \n_+(d\mm) \, \lis g(\mm) = \int_{\L} \m_+(dx)\,g(x) \; ,\qquad
\int \n_{L,k}(d\mm) \, \lis g(\mm) = \int_{\L} \m_{L,k}(dx)\,g(x)\;, $$
where
$$ \lis g(\mm) = \int_0^{\psi(\mm)} dt \, g(\r(\mm,t)) \; , $$
with $\psi(\mm)\in(t_0,\a)$ $\forall \mm\in\MM$.
For any subset $A\subset \MM$, we denote by $\n_+(A)$ the
$\n_+$-measure of $A$: $\n_+(A)=\int_A\n_+(d\mm)$.

We conclude this section with a comment inherited from [CG2].

\*

\\{\bf 2.6.} {\it Remark.} In
Proposition 2.5, we could define the approximating distribution
with $\JJ_{u,k}(x)\to\JJ_{u,k}(x)\,\d_k(x)$, where $\d_k(x)$ $=$
$\sin(\HH_{\TT}^{k/2}x)/\sin(\HH_{\TT}^{-k/2}x)$, which
corresponds to considering a Gibbs state with a different
boundary condition (with the difference becoming irrelevant
in the limit as $k\to\io$, because of the absence of
phase transitions for one-dimensional Gibbs states with short
range interactions, [R2,GL]). Note that the factors $\d_k(x)$ are
{\sl cocycles}, according to [R4], Definition B2.

\vskip1.truecm

\centerline{\titolo 3. Reversible dissipative systems and results.}
\*\numsec=3\numfor=1

\\Let us consider flows $f^t\!\!:M\to M$
verifying the following conditions (A) and (B).

\*

\\{\bf 3.1.} {\cs Definition.} {\it The flow $f^t\!\!:M\to M$
is (A) {\sl dissipative} if
$$ \s_\pm = - \int_\L \m_{\pm}(dx) \, \ln J^{\pm1}(x) > 0 \; , $$
and (B) {\sl reversible} if there is an isometric involution $i\!:M
\to M$, $i^2=\openone$, such that: $if^t=f^{-t}i$.}

\*

If $f^t\!\!:M\to M$ is transitive and reversible, then,
for any $x\in M$, the stable and the unstable manifolds
have the same dimension, so that the the dimension of $M$ is odd.

By reversibility, one has $\s_+=\s_-$, $J(x)=J^{-1}(ix)$,
$iW_x^u=W_{ix}^s$, [G3], \S 2, and $\JJ_{u,k}$ $=$ $\JJ_{s,k}^{-1}(ix)$,
[G3], \S 4. Moreover, if $\L$ is an attractor for
the flow $f^t\!\!: M\to M$, then $\L'=i\L$ is an attractor
for the opposite flow $f^{-t}\!\!:M \to M$,
so that $W_{\L}^s=iW_{\L'}^u$, with the
notations in Proposition 2.4.

\*

\\{\bf 3.2.} {\cs Lemma.} {\it Let $\L$ be an attractor for
the Axiom A flow $ f^t\!\!: M\to M $. If
\acapo
(a) the flow is transitive on $M$, or
\acapo
(b) the flow is reversible and $i\L=\L$,
\acapo
then $\L$ is a connected component of $M$ and
$f^t|\L$ is an Anosov flow.}

\*

\\{\bf 3.3.} {\it Proof of Lemma 3.2.}
If $f^t\!\!: M\to M$ is transitive, $\L$ is dense in $M$
(because of the $f^t$-invariance of $\L$),
and, as $\L$ is closed, then $\L=M$;
this proves (a).\annota{6}{\nota
It is not necessary to assume the existence of an attractor
in order to deduce from transitivity that
$f^t\!\!:M\to M$ is an Anosov flow: in fact transitivity implies
trivially $\O=M$.}

If $\L$ is an attractor, one has
$m(W_{\L}^s)>0$, where $m$ is the measure on $M$ derived from the
Riemann metric, [BR], Theorem 5.6.
If one sets $\L'=i\L$, one has $iW_{\L'}^u=W_{\L}^s$, by reversibility.
If $i\L=\L$, then $m(W_{\L}^u)=m(W_{\L}^s)$.
But $\L=W^u_{\L}$, hence $m(\L)>0$,
so that $\L$ is a connected component of $M$ and
$f^t|\L$ is an Anosov flow, [R5], [BR], Corollary 5.7.
Then (b) follows. \qed

\*

\\{\bf 3.4.} {\cs Remark.}
In hypothesis (b) of Lemma 3.2, one can assume $iW_{\L}^s=W_{\L}^s$
instead of $i\L=\L$: in fact $i\L=\L$ yields $\L=W_{\L}^s$,
and, if $iW_{\L}^s=W_{\L}^s$, one has $i\L\subset iW_{\L}^s= W_{\L}^s$,
hence $i\L=\L$, (because $i\O=\O$).

\*

Note that a stretched exponential bound on the correaltion
functions is obtained, [Ch], for three-dimensional topologically
mixing Anosov flows
satisfying an extra assumption (``uniform nonintegrability''
of the ``foliations'' $E^s$ and $E^u$, [Ch], \S 13, Assumption A5),
while it is known that
topologically mixing Axiom $A$ flows can have correlations function
decaying arbitrarily slowly, [R6,Po].

\*

\\{\bf 3.5.} {\cs Definition.} {\it We define the {\sl dimensionless
volume contraction rate} at $x\in\G(\TT)$ and over a time $k$ as
$$ \e_k(x) ={1\over\s_+k} \sum_{j=-k/2}^{k/2-1}
\ln J^{-1}(\HH_\TT^jx) = {1\over\s_+k}
\ln \JJ^{-1}_k(x) \; , $$
where $\JJ^{-1}_k(x) \= \prod_{j=-k/2}^{k/2-1} J^{-1}(\HH_\TT^jx)$,
and we set $\lis\e_k(\mm)=\int_0^{\psi(\mm)}dt\,\e_k(\r(\mm,t))$.}

\*

Then the following result holds, which can be interpreted as a large
deviation rule, (see [La,CG1]).

\*

\\{\bf 3.6.} {\cs Theorem (Fluctuation theorem).} {\it
Let $\L$ be an attractor for the
dissipative reversible transitive Anosov flow $f^t\!\!:M\to M$.
There exists $p^*>0$ such that the
SRB distribution $\m_+=\m_{\n_+}$ on $\L$ verifies
$$ p-\d\le
\lim_{k\to\io}{1\over\s_+k}\ln{\n_+(\{\mm : \;
\lis\e_k(\mm)\in[p-\d,p +\d]\}) \over
\n_+(\{\mm : \; \lis\e_k(\mm)\in-[p-\d,p+\d]\})} \le p+\d \; , $$
for all $p$ and $\d$ such that $|p|+\d<p^*$.}

\*

If $f^t\!\!:M\to M$ is a dissipative reversible Axiom $A$ flow, and
(1) the restriction of $f^t$ on the attractor $\L$ is
an Anosov flow, and (2)
there exists an isometric involution $i^*\!\!:\L\to\L$, such
that $i^*f^t|\L=f^{-t}i^*|\L$, then Theorem 3.6 still applies.

\vskip1.truecm

\centerline{\titolo 4. Proof of the fluctuation theorem} 
\*\numsec=4\numfor=1

\\For $x\in\G(\TT)$,
the function $\e_k(x)$ can be regarded as a function
on $\MM$, by setting $\lis\e_k(\mm)$ $=$
$\int_0^{\psi(\mm)} dt\,\e_k(\r(\mm,t))$. Then the following
two propositions hold.

\*

\\{\bf 4.1.} {\cs Proposition}. {\it For a suitable $p^*>0$ and
for $p\in(-p^*,p^*)$, $|p|+\d<p_*$, there exists the limit
$$ \lim_{k\to\io} {1\over k} \ln
\n_+(\{\mm : \;\lis\e_k(\mm)\in[p-\d,p+\d]\}) =
\sup_{s\in[p-\d,p+\d]}\big\{ -\z(s) \big\} \; , $$
and $\z(s)$ is a real analytic strictly convex function on
$(-p^*,p^*)$. The difference between the right and left
hand sides tends to zero bounded by $D_1\,k^{-1}$, for
some positive constant $D_1$.}

\*

In Proposition 4.1, $p^*$
is defined as $p^*=\sup_{x\in\L} \{ \limsup_{k\to+\io}
\e_k (\HH_T^{k/2}x)\}$.

\*

\\{\bf 4.2.} {\cs Proposition.} {\it For $\b\in(-\io,\io)$,
define
$$ \l(\b)=\sup_{s\in(-p^*,p^*)}\big\{\b s-\z(s)\big\} \; ; \Eq(4.1) $$
then one has
$$ \l(\b)=\lim_{k\to\io}{1\over k}\ln \int
\n_+(d\mm)\,e^{\b k \lis\e_k(\mm)} \; , $$
and $\l(\b)$ is a real analytic strictly convex function on} $\RR$,
{\it asymptotically linear to $\pm p^*$, for $\b\to\pm\io$.
The difference between the right and left
hand sides tends to zero bounded by $D_2\,k^{-1}$, for
some positive constant $D_2$.}

\*

The two statements are equivalent. In fact Proposition 4.1 yields
Proposition 4.2 (and {\it viceversa}): the proof of such an
assertion is standard, [R2], and is given in in Appendix A2.
Therefore it is enough to prove one of
the two results: the proof of Proposition 4.2 can be deduced from [CO],
and it is reproduced in Appendix A3.

\*

Hence it will be sufficient to prove the following result.

\*

\\{\bf 4.3.} {\cs Lemma.} {\it If $I_{p,\d}=[p-\d,p+\d]$, $|p|+\d<p^*$,
the distribution $\n_+$ verifies the inequalities
$$ {1 \over \s_+k}
\ln { \n_+(\{\mm : \;\lis\e_k(\mm)\in I_{p,\d\mp\h(k)}\}) \over
\n_+(\{\mm : \;\lis\e_k(\mm)\in I_{-p,\d\pm\h(k)})\} }
\cases{<  p+\d+\h'(k)\cr >p-\d-\h'(k)\cr} $$
where $\h(k),\h'(k)>0$ and $\h(k),\h'(k)\to 0$ for $k\to\io$.}

\*

For $q\in\ZZ$ and $n$ odd, set $X=\{q,{q+1},\ldots,{q+n-1}\}
\=[q,q+n-1]$ and
define $\mm_X$ $=$ $(m_q,$ $m_{q+1},$ $\ldots,$ $m_{q+n-1})$ and
$\lis X$ $=$ $q+(n-1)/2$ (the {\sl center} of $X$).
If $\mm\in\MM$, let $\mm_X^0$ be an arbitrary configuration
$\{(m_X^0)_i\}_{i\in\zz}$ such that $(m^0_X)_i$ $=$
$(m_X)_i$, $\forall i=q, \ldots,q+n-1$.

One can write
$$ {1\over\s_+}\ln J^{-1}(\r(\mm,0)) = \sum_{\lis X=0} E_X(\mm_X)
\; , \qquad h_+(\mm) = \sum_{\lis X=0} H_X(\mm_X) \; , \Eq(4.2) $$
where $E_X(\mm_X)$ and $H_X(\mm_X)$ are translation invariant
and exponentially decaying functions, \ie, if
$\th$ denotes translation to the right,
$$ \eqalign{
E_{\th X}(\mm_X) & \= E_X(\mm_X) \; , \qquad
|E_X(\mm_X)| \le b_1^{(E)}\, e^{-b_2^{(E)} n} \; , \cr
H_{\th X}(\mm_X) & \= H_X(\mm_X) \; , \qquad
|H_X(\mm_X)| \le b_1^{(H)}\, e^{-b_2^{(H)} n} \; , \cr} $$
for suitable positive constant $b_1^{(E)}$, $b_2^{(E)}$,
$b_1^{(H)}$ and $b_2^{(H)}$.

Then $k\lis\e_k(\mm)$ can be written as
$$ k\lis\e_k(\mm) \; = \sum_{\lis X \in[-k/2,k/2-1]} E_X(\mm_X) \; , $$
and, if $k\lis\e_k^N(\mm)=\sum^{(N)}E_X(\mm_X)$, with
$\sum^{(N)}$ denoting summation over the sets
$X\subseteq[-k/2-N,k/2+N]$, $N\ge 0$,
while $\lis X \in [-k/2,k/2-1]$, one has the approximation formula
$$ |k\lis\e^N_k(\mm)-k\lis\e_k(\mm)|\le b_1 e^{-b_2 N} \; , $$
where $b_1=(e+1)[(e-1)\,(1-\exp(-b_2^{(E)}))]^{-1}b_1^{(E)}$,
$b_2=b_2^{(E)}$, and $N$ can be chosen $N=0$.
Then
$$ \eqalign{
\n_+ & (\{\mm \; : \;\lis\e^0_k(\mm)\in I_{p,\d-b_1/k}\}) \; \le \;
\n_+(\{\mm \; : \;\lis\e_k(\mm)\in I_{p,\d}\}) \cr
& \le \;
\n_+(\{\mm \; : \;\lis\e^0_k(\mm)\in I_{p,\d+b_1/k}\}) \; . \cr} $$

 From the general theory of one-dimensional Gibbs
states, [R1,R3], (see Proposition 2.5 in \S 2), one has that the
$\n_+$-probability of a configuration $\mm^0_{[-L/2,L/2]}$,
$L\ge k/2$, is
$$ { e^{-{\sum}^* H_X (\mm_X^0) } \; B\left(\mm_{[-L/2,L/2]}\right)
\over \Big[ \sum_{\mm_{[-L/2,L/2]}} e^{-{\sum}^*
H_X (\mm_X^0) } \Big] }  \; , $$
where $\sum^*$ denotes summation over all the $X\subseteq [-L/2,L/2]$,
with $\lis X \in [-k/2,k/2-1]$,
and $B(\mm_{[-L/2,L/2]})$ depends on $\mm_{[-L/2,L/2]}$,
but verifies the bound $|\ln B(\mm_{[-L/2,L/2]})|\le \ln B_2$,
for a suitable $B_2>0$ and uniformly in $L$.

As $\psi(\mm)\in(t_0,\a)$, we can define $\tilde B_2 =
\max\{\a,t_0^{-1}\}$, so that $|\ln \psi(\mm)| \le \ln \tilde B_2$.

Then, for any $L\ge k/2$, one has, for $B_3=B_2\tilde B_2$,
$$ \eqalign{
\n_+ (\{\mm & : \;\lis\e_k(\mm) \in I_{p,\d}\}) 
\; \le \; \n_+(\{\mm : \;\lis \e^0_k(\mm)\in I_{p,\d+b_1/k}\}) \cr
\; \le \; & B_3\,\n_{L,k}(\{\mm : \;\lis\e^0_k(\mm)\in
I_{p,\d+b_1/k}\}) \; \le \; B_3\,
\n_{L,k}(\{\mm : \;\lis\e_k(\mm)\in I_{p,\d+2b_1/k}\}) \; , \cr} $$
and likewise a lower bound is obtained by
replacing $B_3$ by $B_3^{-1}$ and $b_1$ by $-b_1$.

Then, if $I_{p,\d}\subset (-p^*,p^*)$ the set of the
rectangles $T_{\mm_{[-L,L]}}\in\TT_L$ with center $x$
such that $\e_k(x)\in I_{p,\d}$ is not empty, and we have obtained
the following rewriting of Lemma 4.3.

\*

\\{\bf 4.4.} {\cs Lemma.} {\it
The distributions $\n_+$ and $\n_{L,k}$, $L\ge k/2$, verify
the inequalities
$$ {1\over k\s_+}\ln {
\n_+(\{\mm : \;\lis\e_k(\mm)\in I_{p,\d\mp 2b_3/k}\})
\over
\n_+(\{\mm : \;\lis\e_k(\mm)\in- I_{p,\d\pm 2b_3/k}\}) }
\,\cases{
< {1\over k\s_+} \ln B_3^2 + {1\over k\s_+} \ln
{\n_{L,k}(\{\mm : \; \lis\e_k(\mm)\in I_{p,\d}\})
\over
\n_{L,k}(\{\mm : \;\lis\e_k(\mm)\in- I_{p,\d}\}) } \cr
> - {1\over k\s_+} \ln B_3^2 + {1\over k\s_+} \ln
{\n_{L,k}(\{\mm : \; \lis\e_k(\mm)\in I_{p,\d}\})
\over
\n_{L,k}(\{\mm : \; \lis\e_k(\mm)\in- I_{p,\d}\}) } \cr} $$
for $I_{p,\d}\subset (-p^*,p^*)$ and for $k$ so large that
$p+\d+2b_3/k< p^*$.}

\*

Hence Lemma 4.3 follows if the following result can
be proven.

\*

\\{\bf 4.5.} {\cs Lemma.} {\it
There is a constant $\lis b$ such that the approximate
distribution $\m_{L,k}$ verifies the inequalities
$${ 1 \over \s_+k } \ln
{ \n_{L,k}(\{\mm : \;\lis\e_k(\mm)\in I_{p,\d} \}) \over
\n_{L,k}(\{\mm : \;\lis\e_k(\mm)\in-I_{p,\d}\}) } \,
\cases{\le p+\d+ \lis b/k\cr
\ge p-\d -\lis b/k \cr} $$
for $k$ large enough (so that $|p|+\d+\lis b/k<p^*$) and for all
$L\ge k/2$.}

\*

If $\TT$ is a Markov partition also $i\TT$ is such (because $iS=S^{-1}i$
and $i W^u_x=W^s_{ix}$); furthermore if $\TT_1$ and $\TT_2$ are Markov
partitions also $\TT=\TT_1\vee\TT_2$ is such.  Therefore there exists a
time reversal invariant Markov partition $\TT$,
\ie a Markov partition such that $\TT=i\TT$: it is enough to
take any Markov partition $\TT_0$, hence to set
$\TT=\TT_0\vee i\TT_0$.

Since the center of a rectangle $T_{\mm_{-[L,L]}}\in\TT_L$ can be taken
to be any point $x_{\mm_{[-L,L]}}$ in the rectangle $T_{\mm_{[-L,L]}}$
(provided $x_{\mm_{-[L,L]}}=\r(\mm',0)$, $\mm'\in\MM$),
we can and shall suppose that the
centers of the rectangles in $\TT_{\mm_{[-L,L]}}$ have been so chosen
that the center of $i T_{\mm_{[-L,L]}}$ is $i x_{\mm_{[-L,L]}}$, \ie the
time reversal of the center $x_{\mm_{[-L,L]}}$ of $T_{\mm_{[-L,L]}}$.

For $k$ large enough the set of configurations $\mm_{[-L,L]}$
such that $\e_k(x)\in I_{p,\d}$ for all (possible)
$x\in T_{\mm_{[-L,L]}}$ is not empty
and the ratio in Lemma 4.5 can be written, if $x_{\mm_{[-L,L]}}$
is the center of $T_{\mm_{[-L,L]}}$, as
$$ \eqalign{
{ \n_{L,k}(\{\mm : \;\lis\e_k(\mm)\in I_{p,\d} \}) \over
\n_{L,k}(\{\mm : \;\lis\e_k(\mm)\in-I_{p,\d}\}) } & =
{ \sum_{\e_k(x_{\mm_{[-L,L]}})\in I_{p,\d}}
\JJ_{u,k}^{-1}\left(x_{\mm_{[-L,L]}}\right) \over
\sum_{\e_k(x_{\mm_{[-L,L]}})
\in- I_{p,\d}} \JJ^{-1}_{u,k}\left(x_{\mm_{[-L,L]}}\right) } \cr
& =
{ \sum_{\e_k(x_{\mm_{[-L,L]}})\in I_{p,\d}}
\JJ_{u,k}^{-1}\left(x_{\mm_{[-L,L]}}\right) \over
\sum_{\e_k(x_{\mm_{[-L,L]}})
\in I_{p,\d}} \JJ^{-1}_{u,k}\left(i x_{\mm_{[-L,L]}}\right) } \; . \cr} $$
But the time reversal symmetry implies that
$\JJ_{u,k}(x)=\JJ_{s,k}^{-1}(i x)$, so that the above ratio becomes
$$ { \sum_{\e_k(x_{\mm_{[-L,L]}})\in I_{p,\d}}
\JJ_{u,k}^{-1}\left(x_{\mm_{[-L,L]}}\right) \over
\sum_{\e_k(x_{\mm_{[-L,L]}})
\in I_{p,\d}} \JJ_{s,k} \left(x_{\mm_{[-L,L]}}\right) }
\, \cases{< \max_{\mm_{[-L,L]}}
\JJ^{-1}_{u,k}\left(x_{\mm_{[-L,L]}}\right)
\JJ^{-1}_{s,k}\left(x_{\mm_{[-L,L]}}\right) \cr
>\min_{\mm_{[-L,L]}}
\JJ^{-1}_{u,k}\left(x_{\mm_{[-L,L]}}\right)
\JJ^{-1}_{s,k}\left(x_{\mm_{[-L,L]}}\right) \cr} $$
where the maxima are evaluated as $\mm_{[-L,L]}$ varies with
$\e_k(x_{\mm_{-L,L}})\in I_{p,\d}$.

We can replace $\JJ^{-1}_{u,k}(x)\,\JJ^{-1}_{s,k}(x)$ with
$\JJ_k^{-1}(x)B^{\pm1}_4$, $B_4=B_1^3$, and $B_1$ is defined at
the beginning of \S 2.

By definition of the set of $\mm_{[-L,L]}$'s in the maximum operation
in the last inequalities one has
$[\s_+k]^{-1}\ln \JJ^{-1}_k(x_{\mm_{[-L,L]}})
\in I_{p,\d}$: then Lemma 4.5 follows with
$\lis b =\s_+^{-1}\ln B_4$.

 From the chain of implications 4.5 $\to$ 4.4
$\to$ 4.3 $\to$ 3.6, Theorem 3.6 follows and a bound $O(k^{-1})$
is found on the speed at which the limits are approached:
in fact the limit in Lemma 4.1 is reached at speed
$O(k^{-1})$, and the regularity of $\z(s)$, the
size of $\h(k)$ and $\h'(k)$ and the error term in Lemma 4.5
have all order $O(k^{-1})$.

\vskip1.truecm

\0{\bf Acknowledgements.} I would thank G. Gallavotti
for having proposed the argument and for
continuous encouragement and suggestions.
I'm also indebted to M. Cassandro for discussions
about the paper [CO], to F. Bonetto for comments and
remarks, and in particurar to D. Ruelle for
hospitality at IHES, and for enlightening discussions
about the theory of Axiom $A$ systems
and critical comments on the manuscript.

\vskip1.truecm

\centerline{\titolo Appendix A1. Proof of Proposition 2.2}
\*\numapp=1\numfor=1

\\The proof of the statements in Proposition 2.2
can be adapted from [G3].
In fact we can study the map $S=\HH_T\!\!:
\G'(\TT)\to\G'(\TT)$ as it was an Anosov diffeomorphism.
For any $x\in\G(\TT)$, $W_x^s(T)$ and $W_x^u(T)$
are the stable and the unstable manifolds of $x$,
if $T$ is the rectangle in $\G(\TT)$ containing $x$;
the angle $\a(x)$ between them is bounded by two constants
$K_0^{-1}<K_0$, by the transversality implied by the Whitney sum
decomposition (see Definition 1.1).
We can denote by $|DS_{u}(x)|$ and
$|DS_{s}(x)|$ the jacobians of the map $S$ restricted
to the unstable manifold $W_x^u(T)$ and, respectively, to the
stable manifold $W_x^s(T)$.

Then we can apply the discussion
of [G1,G2], and obtain that the measure $\n$ of a cylinder set
$$ \CC\!{\nota \matrix{ -k & \ldots & k \cr m_{-k} & \ldots & m_k \cr} }
= \bigcap_{j=-k}^{k-1} S^{-j}T_{m_j} \subset T_{m_0}$$
is ``essentially'' given by
$$ \n \left( \CC\!{\nota \matrix{ -k & \ldots & k \cr m_{-k}
& \ldots & m_k \cr} } \right)
= \b_{m_{-k}}^s \Big[ \prod_{j=-k}^0 |DS_s(S^jx)| \Big]
\a(x) \Big[ \prod_{j=1}^{k-1} |DS_u^{-1}(S^jx)| \Big] \b_{m_k}^u \; , $$
where $x$ is a point in $\CC\!
{\nota \matrix{ -k & \ldots & k \cr m_{-k} & \ldots & m_k \cr} }$,
$\b_{m_k}^u$ and $\b_{m_{-k}}^s$ denote the
surfaces of the unstable boundary of $T_{m_k}$ and, respectively,
of the stable boundary of $T_{m_{-k}}$. Here ``essentially'' has the
same meaning as in [G2], arising from the approximation
involved by the arbitrarity of the choise of the point $x$,
and it is solved as in [G2].

When the limit as $k\to\io$ is taken, we find that $\n$ is
formally proportional to the exponential of
$$ - \Big[ - \sum_{j=-\io}^{-1} \ln |DS_s(\r(\s^j\mm,0))| - \ln \a(\r(\mm,0))
+ \sum_{j=0}^{\io} \ln |DS_u(\r(\s^j\mm,0))| \Big] \; . $$
By using the fact that $|DS_u|$ and $|DS_s|$ are cocycles,
we can write
$$ \prod_{j=-k}^0 |DS_s(S^jx)| = |DS^k_s(S^{-k}x)| \; , \qquad
\prod_{j=0}^{k-1} |DS_u(S^jx)| = |DS^k_u(x)| \; , $$
and replace $|DS_s^k(S^{-k}x)|$ with
$\JJ_{s,\t_{-k}(x)}(\HH_{\TT}^{-k}x)$, and
$|DS_u^k(x)|$ with $\JJ_{u,\t_{k}(x)}(x)$: the errors caused
by such substitutions are bounded by
two other pairs of constants $K_1^{-1}<K_1$
and $K_2^{-1}<K_2$, by [B2], Lemma 7.1, (this simply means that
the boundary conditions for the corresponding
Gibbs state can be different for finite $k$,
but the difference becomes irrelevant as $k\to\io$; see also
Remark 2.6).

%
%
%
%

Then we have $\m=\n\times m$, where $m$ is the Lebesgue measure
and $\n$ is a measure on $\MM$, so that Proposition 2.2 follows with
$\n_0=\n$ and $B_2=K_0K_1K_2$. \qed

\*

If the volume measure $\m_0$ admits the representation
$\m_{\n_0}=\n_0\times m$, then the measure $\n_+$ describing
the Gibbs state with formal Hamiltonian $H_+(\mm)$
denotes the SRB measure for the map $H_{\TT}\!\!:\G'(\TT)\to\G'(\TT)$,
so that $\m_{\n_+}=\n_+\times m$ is the SRB measure for the $G$.

\vskip1.truecm

\centerline{\titolo Appendix A2. Equivalence of ensembles}
\*\numapp=2\numfor=1

\\In this appendix we prove the equivalence between Proposition
4.1 and Proposition 4.2. Let assume that Proposition 4.1 holds. Define
$$ Q(\b,k) = \int \n_+(d\mm) \, e^{\b k\lis\e_k(\mm)} \; . $$
Given $\d>0$, let $p_0$ be such that $\b p-\z(p) > \l(\b)-c_1\d$
for any $p\in[p_0-\d,p_0+\d]$, for a suitable constant $c_1$. Then
for $k$ large enough (so that $D_1<k\d$)
$$ \eqalign{
Q(\b,k) & \ge \int \n_+(d\mm) \, e^{\b k\lis\e_k(\mm)} \,
\chi (\lis\e_k(\mm)\in[p_0-\d,p_0+\d]) \cr
& \ge \exp \Big[ \b k (p_0 - \d) + k \big( \sup_{s\in[p_0-\d,p_0+\d]}
\big\{ -\z(s) \big\} - k^{-1}D_1 \big) \Big]
\ge e^{k(\l(\b)-\d(2\b+1+c_1))} \; , } $$
(here $\chi$ denotes the characteristic function), so that
$$ \liminf_{k\to\io} {1\over k} \ln Q(\b,k) \ge \l(\b) \; . $$
Given $\d>0$, one can write, for $k$ large enough
(so that $D_1<k\d$),
$$ \eqalign{
Q(\b,k) & = \sum_{i=1}^{k_0-1}
\int \n_+(d\mm) \, e^{\b k\lis\e_k(\mm)} \,
\chi (\e_i\le\lis\e_k(\mm)\le\e_{i+1}) \cr
& + \sum_{i=k_0}^{k-1}
\int \n_+(d\mm) \, e^{\b k\lis\e_k(\mm)}
\, \chi (\e_i\le\lis\e_k(\mm)\le\e_{i+1}) \; , \cr} $$
where the strictly increasing sequence $\{\e_i\}_{i=1}^{k}$ is so taken
that: (a) $-p_*=\e_1<\e_2<\ldots<\e_{k}=p^*$; (b)
$\forall i=1,\ldots,k$, $\e_{i+1}-\e_i\le2\d$;
and (c) $\e_{k_0}=s_0$, where
$\sup_{s\in(-p_*,p_*)}\{-\z(s)\}=-\z(s_0)$. Then
$$ \eqalign{
Q(\b,k) & \le \sum_{i=0}^{k_0-1} e^{\b k\e_{i+1} - k\z(\e_{i+1}) + D_1}
+ \sum_{i=k_0}^{k-1} e^{\b k\e_{i+1} - k\z(\e_{i}) + D_1} \cr
& \le (k-1)\,e^{k[\l(\b)+\d(2\b+1)]} \; , \cr} $$
and one deduces
$$ \limsup_{\k\to\io} {1\over k} \ln Q(\b,k) \le \l(\b) \; , $$
so that the first statement of Proposition 4.2 is proven.
The other properties of $\l(\b)$ follow from the properties
of $\z(s)$, by taking into account that $\l(\b)$ is the
Legendre transform of $\z(s)$.

\vskip1.truecm

\centerline{\titolo Appendix A3. Canonical ensemble}
\*\numapp=3\numfor=1

\\We prove Proposition 4.2. In order to apply the methods of [CO],
let us reformulate it under more general conditions
as follows (recall that real analyticity
means analyticity in an arbitrarily small strip around the real axis).

\*

\\{\bf A3.1.} {\cs Lemma.} {\it Let us denote by $X$ the subsets of
$\ZZ$, by $|X|$ the number of elements in $X$, and define
${\rm diam}(X) = \max_{i,j\in X}|[i,j]|$. Let the
class of the {\rm potentials} be defined as $\F=\{\F_X(\mm_X)\}_{X
\subset \zz}$, where $\F_X(\mm_X)$ is a function depending on the values
of the symbols $\{m_p\}_{p\in X}$, and set
$$ U^{(\F)}(\mm)= \sum_{X\in \zz}\F_X(\mm_X) \; . $$
If $\|f_X\|_{\io}$ denotes the supremum norm for the continuous
function $f_X\in C(\{1,\ldots,\NN\}^{|X|})$,
\ie $\|f_X\|_{\io}=\sup_{\mm_X}|f_X(\mm_X)|$, and
$\BB$ is the Banach space of the {\rm potentials} $\F$
with the norm
$$ \|\F\|_{\BB} \= \sum_{X\ni 0} e^{r\,{\rm diam}(X)}
\|\F_X(\mm_X)\|_{\io} < \io \; , $$
for some $r>0$, then there exists a domain $\o$ of the complex
plain centered in the origin, such that, for $\F\in\BB$, the limit
$$ q(\b) = \lim_{k\to\io} {1\over k} \ln \int \n_{k/2,k}(d\mm)\,
e^{\b U^{(\F)}(\mm) } $$
exists and is analytic in $\b\in\o$.}

\*

Once the existence of the limit $q(\b)$ is proven, the convexity can
be proven with standard methods, [R1], and the linearity in $\b$
for $\b\to\io$ follows from the definition of $p^*$ in \S 4.

\*

\\{\bf A3.2.} {\it Decimation and first cluster expansion.}
Let $\L_p$ be the interval centered at the origin of lenght
$|\L_p|=2pM+(2p+1)L$, and decompose $\L_p$ into consecutive blocks
$A_{-p}$,$B_{-p}$,$A_{-p+1}$,$\ldots$,$A_{p-1}$,$B_{p-1}$,$A_{p}$,
such that $|B_i|=M$ $\forall i=-p,\ldots,p-1$, and $|A_i|=L$
$\forall i=-p,\ldots,p$, with $L\le M$. Set $\ZZ=\lim_{p\to\io}
\L_p$, and define $\G^A_p = \{A_i\}_{i=-p}^{p}$ and
$\G^B_p= \{B_i\}_{i=-p}^{p-1}$.

Consider the function
$$ Z_p(H+\b E) = \sum_{\mm_{\L_p}}e^{-U^{(H+\b E)}(\mm_{\L_p})} \; , $$
where the potentials $H=\{H_X(\mm_X)\}_{X\subset \zz}$
and $E=\{E_X(\mm_X)\}_{X\subset \zz}$
are in $\BB$. Note that $U^{(H+\b E)}=U^{(H)}+U^{(\b E)}$,
so that the real analyticity in $\b$ of
$$ \lim_{p\to\io}|\L_p|^{-1} Z_p(H+\b E) $$
yields Lemma A3.1 (the sign of $\b$ being irrelevant).

If one chooses $H$ and $B$ as defined in \S 4, after Lemma 4.3,
then from Lemma A3.1 Proposition 4.2 follows,
for $A_{ij}\=1$ ($A$ is the matrix introduced in \S 1.8).
The extension to the general case is trivial.

\*

Call $\aa_i=\mm_{A_i}$ and $\bb_i=\mm_{B_i}$ the configurations
in the blocks $A_i$ and $B_i$, and $\aa_S$ [$\bb_S$] the
configuration in $S$, if $S$ is the union
of sets in $\G^A_p$ [$\G^B_p$].
We have
$$ \eqalign{
U^{(H+\b E)}(\mm_{\L_p}) & = \sum_{i=-p}^{p} \a^{(H+\b E)}(\aa_i) +
\sum_{i=-p}^{p-1} J^{(H)}_i(\aa_i, \bb_i, \aa_{i+1}) \cr
& + \sum_{D \in \G^D_p} W^{(H+\b E)}_D(\aa_D) +
\sum_{i=-p}^{p-1} J^{(\b E)}_i(\aa_i, \bb_i,\aa_{i+1}) +
\sum_{C \in \bar\G_p} W^{(H+\b H)}_C (\mm_C)  \; , \cr} $$
where the sets $\G^D_p$ and $\bar \G_p$ are defined as follows:
$$ \G^D_p = \{ D = A_{i_1} \cup \ldots \cup A_{i_k} \; : \;
2 \le k \le p \, , \hbox{ and } D \neq A_i\cup A_{i+1}
\, , \quad \forall i\in\ZZ \} \; , $$
$$ \eqalign{
\bar\G_p = \{ & C = A_{i_1} \cup \ldots \cup A_{i_k} \cup
B_{i_1'} \cup \ldots \cup B_{i_{k'}'} \; : \; 0\le k \le p
\, , \; 1 \le k' \le p+1 \, , \cr
& \hbox{and } C\neq A_i\cup B_i \cup A_{i+1} \, , \;
C\neq A_i B_i \, ,\; C\neq B_i A_{i+1} \, , \;
C \neq B_i \, , \forall i\in\ZZ \} \; , \cr} $$
and
$$ \eqalign{
\a^{(\F)}(\aa_i) & = \sum_{X\subset A_i} \F_X(\aa_X) \; , \cr
J^{(\F)}_i(\aa_i, \bb_i, \aa_{i+1}) & =
\sum_{X \subset A_i\cup B_i \cup A_{i+1} \atop
X \cap B_i \neq \emptyset} \F_X(\mm_X) +
\sum_{X \subset A_i \cup A_{i+1} \atop
X \cap A_i \neq \emptyset \, , \;
X \cap A_{i+1} \neq \emptyset} \F_X(\aa_X) \; , \cr
W^{(\F)}_D (\aa_D) & = \sum_{X\subset D \atop
X \cap A_{i_h} \neq \emptyset \; \forall A_{i_h}
\subset D} \F_X(\aa_D) \; , \cr
W^{(\F)}_C (\mm_C) & = \sum_{X \subset C \atop
X \cap A_{i_h} \neq \emptyset \; \forall A_{i_h} \subset C \, , \;
X \cap B_{i_h'} \neq \emptyset \; \forall B_{i_h'}\subset C}
\F_X(\mm_C) \; . \cr} $$

Note that $\G_p^D=\emptyset$ if only connected subsets $X$ are
allowed for interaction $\F$, as it is the case when $H$ and $E$
are given as in \S 4.

If we define
$$ Z_{B_i}^{(\F)}(\aa_i,\aa_{i+1})
= \sum_{\bb_i} e^{J_i^{(\F)}(\aa_i,\bb_i,\aa_{i+1})} \; , \Eqa(A3.1)$$
and
$$ \eqalign{
\exp[-\tilde U^{(H+\b E)}(\aa_{\G^A_p})] & = \sum_{\bb_{\G^B_p}}
\Big[ \prod_{i=-p}^{p-1} { e^{J^{(H)}_i(\aa_i, \bb_i, \aa_{i+1})}
\over \sum_{\bb_i} e^{ J^{(H)}_i(\aa_i, \bb_i, \aa_{i+1}) } }
\Big] \cdot \cr
& \cdot \Big[ \prod_{C\in \bar\G_p} e^{W^{(H+\b E)}_C (\mm_C) }\Big]
\cdot \Big[ \prod_{i=-p}^{p-1} e^{J^{(\b E)}_i
(\aa_i, \bb_i, \aa_{i+1})} \Big] \; , \cr} $$
then we can average over the variables associated to $\G^B_p$
({\sl decimation procedure}, see [KH1, KH2])
$$ \eqalign{
\sum_{\mm_{\L_p}} U^{(H+\b E)}(\mm_{\L_p}) & =
\sum_{\aa_{\G^A_p}} \Big\{ \sum_{i=-p}^{p} \a^{(H+\b E)}(\aa_i)
+ \sum_{i=-p}^{p-1} \ln Z_{B_i}^{(\F)}(\aa_i,\aa_{i+1}) \cr
& + \sum_{D \in \G^D_p} W^{(H+\b E)}_D(\aa_D) +
\tilde U^{(H+\b E)}(\aa_{\G^A_p}) \Big\} \; . \cr} $$

To each $C \in \bar\G_p$ we associate a bond $\p(C)$, and to each
$B\in\G^B_p$ a bond $\p(B)$, such that the lenght of a bond $\p(B)$
is $|\p(B)|=1$ and the lenght of a bond $\p(C)$, denoted as $|\p(C)|$,
is given by the number of blocks $A$'s and $B$'s contained in $C$.
Consider
$$ \RRR = \{ C_1 , \ldots , C_k , B_1 , \ldots , B_h \Big\}
\; , \qquad C_{s} \subset \bar\G_p \; \forall s=1,\ldots,k \; , $$
and set $\tilde C = C \cap \G^B_p$ and $\tilde \RRR =
\{ \tilde C_1 , \ldots , \tilde C_k , B_1 , \ldots , B_h \Big\}$:
$|\p(\tilde C)|$ is given by the number of sets $B$'s contained in $C$.
We set $\tilde B=B$.

The set of bonds corresponding to $\RRR$, \ie
$$ R = \Big\{ \p(C_1), \ldots, \p(C_k), \p(B_1) ,
\ldots , \p(B_h) \Big\} \; , \qquad C_{s} \subset \bar\G
 \; \forall s=1,\ldots,k \; , \Eqa(A3.2) $$
is a {\sl polymer} (see [GMM]) if, for any choise of bonds $\p(X_l)$
and $\p(X_j)$, with $X_l, X_j\in\RRR$, there exist $X_{i_1},
\ldots,X_{i_r}\in\RRR$, $r\le k+h$, such that
$X_{i_1}=X_l$, $X_{i_r}=X_j$ and $\tilde X_{i_h} \cap \tilde X_{i_h+1}
\neq \emptyset$ $\forall h=1,\ldots,r-1$. Then one has
$$ \exp[-\tilde U^{(H+\b E)}(\aa_{\G^A_p})] = 1 + \sum_{n=1}^{\io}
\sum_{R_1,\ldots,R_n
\atop \tilde R_i \cap \tilde R_j = \emptyset}
\prod_{i=1}^n \z (R_i) \; , $$
where $\tilde R$ is defined as $R$ in \equ(A3.2),
but with $C_s$ replaced with $\tilde C_s$ $\forall s=1,\ldots,k$, and
$$ \eqalign{
\z(R) & = \sum_{\bb_{\tilde \RRR}}
\Big[ \prod_{B_i \subset \tilde \RRR}
{e^{J^{(H)}_i(\aa_i, \bb_i, \aa_{i+1})}
\over \sum_{\bb_i} e^{ J^{(H)}_i(\aa_i, \bb_i, \aa_{i+1}) } }
\Big] \cdot \cr
& \cdot \Big[ \prod_{s=1}^{k} \Big(
e^{W^{(H+\b E)}_{C_s} (\mm_{C_s}) } - 1 \Big) \Big]
\cdot \Big[ \prod_{s'=1}^{h} \Big( e^{J^{\b E}_{i_{s'}}
(\aa_{i_{s'}}, \bb_{i_{s'}}, \aa_{i_{s'}+1})}
- 1 \Big) \Big] \; , \cr} $$
is the {\sl activity} of the polymer $R$. Here $B_i\subset \tilde\RRR$
means $B_i\in\RRR$ or $B_i\subset \tilde C_j\in\tilde\RRR$ for some
$j=1,\ldots,k$. We set $|\tilde R|$
$=$ $\sum_{s=1}^k|\p(\tilde C)|$ $+h$ (as $|\p(B)|=1$).

\*

\\{\bf A3.3.} {\cs Lemma.} {\it Given a potential $\Phi\in\BB$, and
considered a polymer $R$, the activity $\z(R)$ satisfies the inequality
$$ |\z(R)| \le \r^{|\tilde R|} \prod_{s=1}^k w_{C_s} \prod_{s'=1}^h
j_{B_{s'}} \; , $$
where $\r=e^{-r}$, $w_C = 2\,e^{r|\p(\tilde C)|}
\| W_C^{(H+\b E)}\|_{\io}$ and
$j_{B_i}=2\,e^r$ $\|J_i^{(\b E)}\|_{\io}$, being $\|f_X\|_{\io}$
the supremum norm for the continuous function $f_X$, and
$w_C$ and $j_B$ are positive constants such that
$\max\{w_C,j_B\} \le \ln [\sqrt{\r}(2-\sqrt{\r})]^{-1}$
for $\b$ small enough and $L$ sufficiently large.}

\*

\\{\bf A3.4.} {\it Proof of Lemma A3.3.} For complex $z$ such that
$|z|<1/2$, one has $|e^z-1|\le 2|z|$. Since
$$ \lim_{L\to\io} e^{r'L} \, \| W_C^{(H+\b E)} \|_{\io} = 0 \; ,
\qquad \forall r'<r \; , $$
and
$$ \lim_{\b\to 0} \b^{1-\e}\,\|J_i^{(\b E)}\|_{\io} = 0 \; , 
\qquad \forall \e>0 \; , $$
we can apply the above inequality, and obtain
$$ |\z(R)| \le \sum_{\bb_{\tilde R}}
\Big[ \prod_{B_i\subset \tilde R}
{e^{J^{(H)}_i(\aa_i, \bb_i, \aa_{i+1})}
\over \sum_{\bb_i} e^{J^{(H)}_i(\aa_i, \bb_i, \aa_{i+1})} } \Big] \cdot
\Big( \prod_{s=1}^k 2 \| W_C^{(H+\b E)} \| \Big) \cdot
\Big( \prod_{s'=1}^h 2 \|J_{j_{s'}}^{(\b E)}\| \Big) \; ; $$
then we can
\acapo
\\(1) extract a factor $e^{-r|\p(\tilde C)|}$ from
$\| W_C^{(H+\b E)} \|_{\io}$ and a factor $e^{-r}$ from
$\|J_{j_{s'}}^{(\b E)}\|_{\io}$, because of the exponential decay
of the interaction, and\\
\acapo
\\(2) make $w_C$ and $j_B$ arbitrarily small
by taking $L$ sufficiently large, and
$$ \eqalign{
& L > {1\over r} \ln \Big[ 4 (1-\r)^{-1} \|H+\b E\|_{\BB}
\max\{1, \ln [\sqrt{\r} (2-\sqrt{\r})] \} \Big] \; , \cr
& |\b| < {\r \over 4M\|E\|_{\BB} } \min\{ 1,\ln[\sqrt{\r}
(2-\sqrt{\r})]^{-1}\} \; , \cr} $$
so obtaining Lemma A3.3. \qed

\*

\\{\bf A3.5.} {\cs Remark.}
Note that, if we had considered an interaction
with exponential decay $e^{-r|X|}$ instead of $e^{-r\,{\rm diam}(X)}$,
the same bound as in Lemma A2.3 would have followed. In fact
the same cluster expansion can be still performed, and the only
difference is that now $W_C^{(H+\b E)}$ decays as $e^{-r|\tilde C|}$:
but this is sufficient in order to prove Lemma A3.3.

\*

\\{\bf A3.6.} {\it Second cluster expansion.}
By Lemma A3.3, for suitable $\b$ and $L$, one has
$$ |\z(R)| \le \r^{|\tilde R|} \prod_{C\in\RRR} \CC_C \; ,
\qquad 0 < \CC_C \le K \; , \qquad K \le \ln[\sqrt{\r}
(2-\sqrt{\r})]^{-1} \; ; $$
then the conditions of [CO], Lemma 1,
are satisfied,  so that we can deduce (analogously to [CO], Lemma 2)
$$ \exp[ -\tilde U^{(H+\b E)}(\aa_{\G^A_p})] = \sum_{D\in \tilde\G^D_p}
\tilde W^{(H+\b E)}_D(\aa_D) \; , $$
where $\tilde\G^D_p$ is defined as $\G^D_p$, but with no
restriction, and
$\tilde W^{(H+\b E)}_D(\aa_D)$ is analytic in $\b\in\o_M$,
if $\o_M$ is a circle around the origin of the complex plane
whose radius tend to zero as $M\to\io$. One can write
$$ \tilde W^{(H+\b E)}_D(\aa_D)
= \sum_{R_1,\ldots,R_n \atop \cup_{i=1}^n
\RRR_i\setminus \tilde \RRR_i \subseteq D
} \varphi_T(R_1,\ldots,R_n) \prod_{i=1}^n
\z(R_i) \; , $$
where the sum is over all the polymers $R_1,\ldots,R_n$
such that the product $\z(R_1)$ $\ldots$ $\z(R_n)$ depends only on the
variables $\aa_D$, and $\varphi_T$ $(R_1,$ $\ldots,$ $R_n)$ is a suitable
coefficient, (see [CO], Lemma 1; see also [GMM]).

Then we can define (recall \equ(A3.1) and define ${\bf 1}$ as the
configuration of a block $A$ with each element set equal to $1$)
$$ \bar\a^{(H)}(\aa_i) = \a^{(H)}(\aa_i) + \ln
\Big[ { Z_{B_i}^{(H)}(\aa_i,{\bf 1}) \cdot
Z_{B_{i-1}}^{(H)}({\bf 1},\aa_i) \over  
Z_{B_i}^{(H)}({\bf 1},{\bf 1}) \cdot
Z_{B_{i-1}}^{(H)}({\bf 1},{\bf 1}) } \Big] \; , $$
and
$$ \VV_D (\aa_D) = \cases{
\a_i^{(\b E)}(\aa_i) \; , & if $D=A_i \;$, \cr
\tilde W_{A_i\cup A_{i+1}}^{(H+\b E)}(\aa_i,\aa_{i+1}) +
\ln
\Big[ { Z_{B_i}^{(H)}(\aa_i,\aa_{i+1}) \cdot
Z_{B_{i}}^{(H)}({\bf 1},{\bf 1}) \over  
Z_{B_i}^{(H)}(\aa_i,{\bf 1}) \cdot
Z_{B_{i}}^{(H)}({\bf 1},\aa_{i+1}) } \Big]  \; , &
if $D=A_i\cup A_{i+1} \;$, \cr
W_D^{(H+\b E)}(\aa_D) + \tilde W^{(H+\b E)}_D(\aa_D)
\; , & if $D \neq A_i, A_i\cup A_{i+1} \;$, \cr} $$
and write
$$ \eqalign{
\sum_{\mm_{\L_p}} \exp[ -U^{(H+\b E)}(\mm_{\L_p}) ] & =
\hbox{const.} \Big[ \prod_{i=-p}^{p} \sum_{\aa_i}
e^{ \bar \a^{(H)}(\aa_i) } \Big] \cdot \cr
& \cdot \sum_{\aa_{\G^A_p}} \Big[ \prod_{i=-p}^{p}
{ e^{ \bar \a^{(H)}(\aa_i) } \over \sum_{\aa_i}
e^{ \bar \a^{(H)}(\aa_i) } } \Big] \cdot \Big[
\prod_{D\in\tilde\G^D_p} e^{\VV_D(\aa_D)} \Big] \; . \cr} $$

We introduce a new cluster expansion by associating to each $D \in
\tilde\G^D_p$ a bond $\p(D)$, and defining a polymer $S$ as
$$ S = \{ \p(D_1) , \ldots , \p(D_k) \} \; , \Eqa(A3.3) $$
and $\SSS = \{D_1,\ldots,D_k\}$. Then
$$ \sum_{\mm_{\L_p}} \exp[-U^{(H+\b E)}(\mm_{\L_p}) ] =
\Big[ \prod_{i=-p}^{p} \sum_{\aa_i} e^{ \bar \a^{(H)}(\aa_i) }
\Big] \cdot
\Big( 1 + \sum_{n=1}^{\io} \sum_{S_1,\ldots,S_n
\atop S_i \cap S_j = \emptyset} \prod_{i=1}^n
\Theta(S_i) \Big) \; , $$
where
$$ \Theta(S) = \sum_{\aa_\SSS} \prod_{A_i \subset \SSS} 
{ e^{ \bar \a^{(H)}(\aa_i) } \over \sum_{\aa_i} e^{ \bar \a^{(H)}(\aa_i) } }
\prod_{D\in \SSS} \Big( e^{\VV_D(\aa_D)} - 1 \Big) $$
is the activity of the polymer $S$. Here $A_i\subset\SSS$
means $A_i\subset D_j\in\SSS$ for some $j=1,\ldots,k$. 

\*

\\{\bf A3.7.} {\cs Lemma.} {\it If
$Z_{B_{i}}^{(H)}(\aa_i,\aa_{i+1})$ is defined as
$Z_{B_{i}}^{(H)}(\aa_i,\aa_{i+1})=\sum_{\bb_{i}}\exp J_i^{(H)}
(\aa_i,\bb_i,\aa_{i+1})$, then
$$ \lim_{M\to\io} \sup_{\aa_i,\aa_{i+1}} \Big\{ \ln
\Big[ { Z_{B_i}^{(H)}(\aa_i,\aa_{i+1}) \cdot
Z_{B_i}^{(H)}({\bf 1},{\bf 1}) \over  
Z_{B_i}^{(H)}(\aa_i,{\bf 1}) \cdot Z_{B_i}^{(H)}({\bf 1},\aa_{i+1}) }
\Big] \Big\} = 0 \; , $$
$\forall L<\io$.}

\*

\\{\bf A3.8.} {\it Proof of Lemma A3.7.}
Let $\ZZ^+=\{i\in\ZZ\;:\;i\ge 0\}$ and $K_+=\{1,\ldots,\NN\}^{\zz^+}$.
We denote by $C(K_+)$ the Banach space of the real continuous functions
on $K_+$, and by $M^*(K_+)$ its dual, \ie the space of
real measures on $K_+$. Given a
configuration $\mm_N\in \{1,\ldots,\NN\}^{N}$, $N\ge1$,
and a configuration $\mm_+\in K_+$, one can define the configuration
$(\mm_N,\mm_+)\in K_+$ as
$$ (\mm_N,\mm_+)_i = \cases{ (\mm_N)_i \; , & for $i=0,\ldots,N \;$, \cr
(\mm_+)_{i-N} \; , &  for $i>N$. \cr} $$

Given $\F\in\BB$, an operator $\LL_{\F}\!\!:C(K_+)\to C(K_+)$
is defined by
$$ \LL_{\F}f\,(\mm_+) = \sum_{m_0=1}^{\NN} \exp\Big[
\sum_{X \subset \zz^+ \atop X \ni 0}
\F_X(\mm_{X}) \Big] f(m_0,\mm_+) \; ; $$
then there exist $\l_{\F}>0$, $h_{\F}\in C(K_+)$ and
$\n_{\F}\in M^*(K_+)$ such that
\acapo
(a) $\LL_{\F}h_{\F}=\l_{\F}h_{\F}$, and
\acapo
(b) if $f\in C(K_+)$,
$\lim_{k\to\io} \| \l_{\F}^{-k}\LL^k_{\F}f-\n_{\F}(f)\,h_{\F}
\|_{\io} = 0$, uniformly for $\F$ in a bounded subset of a finite dimensional
subspace of $\BB$.

The proof of such a statement follow from [R1] and
can be found in [G1], Ch. 18, Proposition XXXV and exercises,
(it is essentially an adaptation from [R1], see also [GL]).

Define the function in $C(K_+)$
$$ f_{\aa_i}(\mm_+)=\exp\Big[ \sum_{X\subset A_i\cup A_{i+1}}
\F_X(\mm_X) \Big] \; , $$
where $\aa_i\in\{1,\ldots,\NN\}^L$. Note that
$f_{\aa_i}(\mm_+)$ depends only on the first $L$
symbols of $\mm_+$, (so that the successive
ones can be set equal to an arbitrary value, say 1). Then one has
$$ { Z_{B_i}^{(H)}(\aa_i,\aa_{i+1}) \cdot
Z_{B_i}^{(H)}({\bf 1},{\bf 1}) \over  
Z_{B_i}^{(H)}(\aa_i,{\bf 1}) \cdot Z_{B_i}^{(H)}({\bf 1},\aa_{i+1}) }
= { \LL_{\F}^M f_{\aa_i}(\aa_{i+1},{\bf 1})
\LL_{\F}^M f_{\bf 1}({\bf 1}) \over
\LL_{\F}^M f_{\aa_i}({\bf 1})
\LL_{\F}^M f_{\bf 1}(\aa_{i+1},{\bf 1}) } \; , $$
where ${\bf 1}$ appearing in $(\aa_{i+1},{\bf 1})$ is an element
in $K_+$, while the subscript ${\bf 1}$ in $f_{\bf 1}$ is an element in
$\{1,\ldots,\NN\}^L$ (\ie $\aa_i={\bf 1}$).

Then from the property (b) above, Lemma A3.7 follows. \qed

\*

\\{\bf A3.9.} {\it Proof of Lemma A3.1.}
We consider the cluster expansion envisaged in \S A3.5. From the
interaction $\{\VV_D\}_{D\in\tilde\G^D_p}$,
terms of the following form arise:
\acapo
\\(a) $\sum_{R_1,\ldots,R_n}\varphi_T(R_1,\ldots,R_n)
\, \z(R_1)\ldots\z(R_n)$, where the dependence on a
configuration $\aa_i$ is only through the factors
$$ U_i={e^{J_i^{(H)}(\aa_i,\bb_i,\aa_{i+1})} \over
\sum_{\bb_i} e^{J_i^{(H)}(\aa_i,\bb_i,\aa_{i+1})} } $$
appearing in $\z(R)$;
\acapo
\\(b) for $D=A_i$, $\a_i^{(\b E)}(\aa_i)$;
\acapo
\\(c) $\sum_{R_1,\ldots,R_n}\varphi_T(R_1,\ldots,R_n)
\, \z(R_1)\ldots\z(R_n)$, where the dependence on the
configurations $\aa_i$ is (also) through terms
$W_C^{(H+\b E)}(\mm_C)$
and $J_i^{(\b E)}(\aa_i,\bb_i,\aa_{i+1})$ in $\z(R)$;
\acapo
\\(d) for $D=A_i\cup A_{i+1}$, also
$$ \ln \Big[ { Z_{B_{i}}^{(H)}(\aa_i,\aa_{i+1}) \cdot 
Z_{B_{i}}^{(H)}({\bf 1},{\bf 1}) \over  
Z_{B_{i}}^{(H)}(\aa_i,{\bf 1}) \cdot
Z_{B_{i}}^{(H)}({\bf 1},\aa_{i+1}) } \Big] \; . $$

As a polymer $R$ contains at most $2|\tilde R|$
$A$-blocks through the factors $U_i$, we can extract also
a factor $\r^{1/4}$ from each $A$-block appearing in terms
of the form (a), by simply replacing $\z(R)$ with a new
activity $\hat \z(R)=\r^{-|\tilde R|/2}\r(R)$.

In (b), we can write $\a_i^{(\b E)}(\aa_i)=\r^{1/2}[\r^{-1/2}
\a_i^{(\b E)}(\aa_i)]$, where $\r^{-1/2}\|\a_i^{(\b E)}\|_{\io}$ can
be made arbitrarily small by taking $\b$ small enough.

As far as the terms in (c) are concerned, we can extract
a factor $e^{-r}=\r$ from each $A$-block, thanks to the
exponential decay of the interaction, and the remaining
factor can be made arbitrarily small by taking $L$ large
and $\b$ small, (see the proof of Lemma A3.4 and the
definition of the set $\G_p^D$ in \S A3.2).

By Lemma A3.7, we can extract a factor $\r$ from each term
in (d) by taking $M$ sufficiently large.

\*

Therefore we have a factor $\r^{1/4}$ $\forall A$ arising in
(a) and a factor $\r^{1/2}$ $\forall A$ arising in (b), (c) and (d).
Then we can bound
$$ |\Theta(S)| \le \tilde\r^{|S|}\prod_{D\in \SSS} \CC_D \; , $$
where $\tilde\r =\r^{1/4}$, and
$$ \CC_D = \cases{ 2\r^{-1/2} \|\a^{(\b E)} \|_{\io} \; ,
& if $D=A_i \; $, \cr
2\r^{-1} \|\VV_D\|_{\io} \; , & if $D=A_i\cup A_{i+1} \; $, \cr
2 \big[ \|W_D\| + \|\hat W_D\|_{\io} \big] \; , &
if $D\neq A_i\,,\,A_i\cup A_{i+1} \;$
, \cr} $$
being $\hat W_D$ defined as $\tilde W_D$, but with $\hat \z(R)$
replacing $\z(R)$. 

If $\b\to 0$ and $M\to \io$, then
$$ \sum_X |\VV_X(\aa_X)|\,e^{|X|r'} $$
can me made arbitrarily small, for some $r'<r$,
in order to apply Israel's analyticity
theorem, [Is], Theorem II, 4, (see also [CO]).
Equivalently we can reason as before, and we can apply
again [CO], Lemma 1, and deduce that, for any constant $\bar\k>0$,
we have
\acapo
\\(1) for any $L$, $\exists M_1(L)$ such that for $\forall M\ge M_1(L)$
$$ {2\over \r} \sup_{\aa_i,\aa_{i+1}}
\Big\{ \ln \Big[ { Z_{B_{i}}^{(H)}(\aa_i,\aa_{i+1}) \cdot 
Z_{B_{i}}^{(H)}({\bf 1},{\bf 1}) \over  
Z_{B_{i}}^{(H)}(\aa_i,{\bf 1}) \cdot
Z_{B_{i}}^{(H)}({\bf 1},\aa_{i+1}) } \Big]
\Big\} \le {\bar\k \over 3} \; ; $$
\acapo
\\(2) $\exists L_0$ such that $\forall L\ge L_0$, $\forall M$ and
$\forall \b\in\o_M$ (being $\o_M$ defined in \S A3.6)
$$ \sup_{A\in \G^A_p} \sum_{D\supset A} 2 \| \hat W_D(\aa_D) \|_{\io}
\le {\bar\k \over 3} \; ; $$
\acapo
\\(3) for $L=L_0$, $\exists M_2(L_0)$ such that
$\forall \b \in \o_{M_2(L_0)}$ and $\forall M\ge M_2(L_0)$
$$ \sup_{i} {2\over\sqrt{\r}} \|\a_i^{(\b E)} \|_{\io} +
\sup_{A\in \G^A_p} \sum_{D\supset A} 2 \| W_D(\aa_D) \|_{\io}
\le {\bar\k \over 3} \; . $$

Therefore, if $\b\in \o_{M_0}$, with $M_0=\max\{M_1(L_0),M_2(L_0)\}$,
there exist $\k\=\k(\b,L_0,M_0)\le \bar\k$ such that
$\CC_D\le\k$. If $\bar\k$ is so chosen that
$\k$ $\le$ $\ln[\sqrt{\tilde\r}$ $(2-\sqrt{\tilde\r})]^{-1}$,
one can apply again [CO], Lemma 1: then 
there is a constant $G(\tilde\r,\k)$ such that 
$$ \lim_{p\to\io}{1\over|\L_p|} \ln Z_p(H+\b E) \le
G(\tilde\r,\k) + {1\over L} \ln \Big| \sum_{\aa_i} e^{\bar\a^{(H)}(\aa_i)}
\Big| \; . $$
The uniformity of the bound and the existence of the
second limit uniformly in $L$ (for real $H$, the proof of
such a result is standard, [R2]) allow us to apply Vitali's
convergence theorem, [T], \S 5.21,
and complete the proof of Lemma A3.1. \qed

\vskip1.truecm

\centerline{\titolo References}
\*

\halign{\hbox to 1.2truecm {[#]\hss} & 
        \vtop{\advance\hsize by -1.25 truecm \\#}\cr

A& {D.V. Anosov:
{\sl Geodesic flows on closed Riemann manifolds with negative
curvature}, Proc. Steklov Inst. Math. {\bf 90} (1967),
American Mathematical Journal, Providence, Rhode Island, 1969. }\cr
%
%
Bb& {N. Bourbaki:
{\sl El\'ements de Math\'ematique, Topologie G\'en\'erale},
Hermann, Paris, 1963. }\cr
B1& {R. Bowen:
One-dimensional hyperbolic sets for flows,
{\it J. Differential Equations} {\bf 12}, 173--179 (1972). }\cr
B2& {R. Bowen:
Symbolic dynamics for hyperbolic flows.
{\it Amer. J. Math.} {\bf 95}, 429--459 (1973). }\cr
B3& {R. Bowen:
{\sl Equilibrium states and the ergodic theory of Anosov
diffeomorphisms}, Lectures Notes on Math. {\bf 470},
Springer, Berlin, 1975. }\cr
B4& {R. Bowen:
{\sl On Axiom $A$ diffeomorphisms}, Regional Conference Series in
Mathematics {\bf 35}, American Mathematical Society, Providence,
Rhode Island, 1978. }\cr
BR& {R. Bowen, D. Ruelle:
The ergodic theory of Axiom $A$ flows.
{\it Inventiones Math.} {\bf 29}, 181--202 (1975). }\cr
BW& {R. Bowen, P. Walters:
Expansive one-parameter flows,
{\it J. Differential Equations} {\bf 12}, 180--193 (1972). }\cr
Ch& {N.I. Chernov:
Markov approximations and decay of correlations for Anosov flows,
Preprint (1995). }\cr
CG1& {E.G.D. Cohen, G. Gallavotti:
Dynamical ensembles in stationary states,
{\it J. Stat. Phys.} {\bf 80}, 931--970 (1995). }\cr
CG2& {E.G.D. Cohen, G. Gallavotti:
Chaoticity hypothesis and Onsager re\-ci\-pro\-ci\-ty,
Pre\-print (1995). }\cr
CO& {M. Cassandro, E. Olivieri:
Renormalization group and analyticity in one dimension.
A proof of Dobrushin's theorem,
{\it Comm. Math. Phys.} {\bf 80}, 255--269 (1981). }\cr
ER& {J.-P. Eckmann, D. Ruelle:
Ergodic theory of chaos and strange attractors,
{\it Rev. Mod. Phys} {\bf 57}, 617--656 (1985). }\cr
%
%
FW& {J. Franks, R. Williams: Anomalous Anosov flows,
in {\sl Global theory of dynamical systems}, Proceedings, Evanston (1979),
Lecture notes in Math. {\bf 819}, 158--174, Ed. Z. Nitecki \&
C. Robinson, Springer, Berlin, 1980. }\cr
G1& {G. Gallavotti:
{\sl Aspetti della teoria ergodica, qualitativa
e sta\-ti\-sti\-ca del mo\-to}, Pi\-ta\-go\-ra, Bo\-lo\-gna, 1981. }\cr
G2& {G. Gallavotti:
Topics on chaotic dynamics, in {\sl Third Granada Lectures in
Computational Physics. Proceedings, Granada, Spain, 1994},
Lectures Notes on Phys. {\bf 448}, Ed. P.L. Garrido \&
J. Marro, Springer, 1995. }\cr
G3& {G. Gallavotti:
Reversible Anosov diffeomorphisms and large deviations, 
{\it Math. Phys. Electronic J.} {\bf 1}, 1--12 (1995). }\cr
GL& {G. Gallavotti, T.F. Lin:
One dimensional lattice gases with rapidly decreasing interactions,
{\it Arch. Rational Mech. Anal.} {\bf 37}, 181-191 (1970). }\cr
GMM& {G. Gallavotti, A. Martin-L\"of, S. Miracle-Sole:
Some problems connected with the description of coexisting phases
at low temperature in the Ising model, in {\sl Statistical
mechanics and mathematical problems}, Battelle Seattle Rencontres (1971)
Lectures Notes on Phys. {\bf 20}, 162--203,
Ed. A. Lenard, Springer, Heidelberg, 1973. }\cr
KH1& {L.P. Kadanoff, A. Houghton:
Numerical evaluations of the critical properties of the
two--dimensional Ising model,
{\it Phys. Rev} {\bf B11}, 377--386 (1975). }\cr
KH2& {L.P. Kadanoff, A. Houghton:
in {\sl Renormalization group in critical phenomena and
quantum field theory}, Proceedings of the Temple University Conference
on critical phenomena, 1973,
Ed. J.D. Gunton \& M.S. Green, Dept. of Physics, Temple University,
Phyladelphia, 1973. }\cr
%
%
HP& {M. Hirsch, C.C. Plug:
Stable manifolds and hyperbolic sets, in {\sl Global Analysis},
{\it Proc. Symp. in Pure Math.} {\bf 14}, 133--163 (1970). }\cr
Is& {R.B. Israel:
High temperature analyticity in classical lattice systems,
{\it Comm. Math. Phys.} {\bf 50}, 245--257 (1976). }\cr
La& {O. Lanford: Entropy and equilibrium states in classical
statistical mechanics, in {\sl Statistical
mechanics and mathematical problems}, Battelle Seattle Rencontres (1971)
Lectures Notes on Phys. {\bf 20}, 1--113,
Ed. A. Lenard, Springer, Heidelberg, 1973. }\cr
%
%
%
Pl& {J. F. Plante:
Anosov flows, {\it Amer. J. Math.} {\bf 94}, 729--754 (1972). }\cr
PS& {C.C. Plug, M. Shub:
The $\O$-stability theorem for flows,
{\it Inventiones Math.} {\bf 11}, 150--158 (1970). }\cr
Po& {M. Pollicott:
On the rate of mixing of Axiom $A$ flows,
{\it Invent. Math.} {\bf 81}, 413--426 (1985). }\cr
R1& {D. Ruelle:
Statistical mechanics of a one dimensional lattice gas,
{\it Comm. in Math. Phys.} {\bf 9}, 267--278 (1968). }\cr
R2& {D. Ruelle:
{\sl Statistical Mechanics}, Benjamin, New York, 1969. }\cr
R3& {D. Ruelle:
{\sl Thermodynamics Formalism},
Encyclopedia of Mathematics Vol. 5, Addison--Wesley, Reading, 1978. }\cr
R4& {D. Ruelle:
Ergodic theory of differentiable dynamical systems,
{\it Pu\-bli\-ca\-tions Ma\-th\'e\-ma\-ti\-ques de l'IHES}
{\bf 50}, 275--306 (1979). }\cr
R5& {D. Ruelle:
Measures describing a turbulent flow,
{\it Ann. N.Y. Acad. Sc.} {\bf 357}, 1--9 (1980). }\cr
R6& {D. Ruelle:
Flots qui ne m\'elangent pas exponentiellement,
{\it C. R. Acad. Sc. Paris} {296}, 191--193 (1983). }\cr
Si1& {Ya. G. Sinai: Markov partitions and $C$-diffeomorphisms,
{\it Func. Anal. and its Appl.} {\bf 2}, 61--82 (1968). }\cr
Si2& {Ya. G. Sinai: Construction of Markov partitions, {\it Func. Anal.
and its Appl.} {\bf 2}, 245--253 (1968). }\cr
Si3& {Ya. G. Sinai: Gibbs measures in ergodic theory, {\it Russ.
Math. Surveys} {\bf 27}, No. 4, 21--64 (1972). }\cr
Sm& {S. Smale: Differentiable dynamical systems,
{\it Bull. A. M. S.} {\bf 73}, 747--817 (1967). }\cr
T& {E.C. Titchmarsch: {\sl The theory of functions},
Oxford University Press, London, 1933; second edition, 1939. }\cr
}

\ciao